\documentclass[%
 aip,
 jcp, % or other AIP journal style
 nofootinbib,
 amsmath,amssymb,
 reprint,
]{revtex4-1}

\usepackage{graphicx}% Include figure files
\usepackage{dcolumn}% Align table columns on decimal point
\usepackage{xcolor}
\usepackage{adjustbox}
\usepackage{import}
\usepackage{booktabs}
\usepackage{mathtools}
\usepackage{comment}
\usepackage{hyperref}
\usepackage{siunitx}   % For decimal alignment
\DeclarePairedDelimiter\bra{\langle}{\rvert}
\DeclarePairedDelimiter\ket{\lvert}{\rangle}
\DeclarePairedDelimiterX\braket[2]{\langle}{\rangle}{#1\,\delimsize\vert\,\mathopen{}#2}

\usepackage{bm}% bold math

\usepackage[utf8]{inputenc}
\usepackage[T1]{fontenc}
\usepackage{mathptmx}
\usepackage{etoolbox}

\newcommand\blfootnote[1]{%
  \begingroup
  \renewcommand\thefootnote{}\footnote{#1}%
  \addtocounter{footnote}{-1}%
  \endgroup
}

\DeclareUnicodeCharacter{2500}{\textemdash} 
\DeclareUnicodeCharacter{2212}{-}

\makeatletter
\def\@email#1#2{%
 \endgroup
 \patchcmd{\titleblock@produce}
  {\frontmatter@RRAPformat}
  {\frontmatter@RRAPformat{\produce@RRAP{*#1\href{mailto:#2}{#2}}}\frontmatter@RRAPformat}
  {}{}
}%
\makeatother
\begin{document}

\preprint{AIP/123-QED}

\title{Method--independent cusps for atomic orbitals in quantum Monte Carlo}

% Force line breaks with \\
\author{Trine Kay Quady${}^\dagger$}
 \affiliation{ 
Department of Chemistry, University of California, Berkeley, California 94720, USA%\\This line break forced with \textbackslash\textbackslash
}
\author{Sonja Bumann${}^\dagger$}%

\affiliation{ 
Department of Chemistry, University of California, Berkeley, California 94720, USA%\\This line break forced with \textbackslash\textbackslash
}%
\affiliation{%
Chemical Sciences Division, Lawrence Berkeley National Laboratory, Berkeley, California 94720, USA%\\This line break forced% with \\
}

\author{Eric Neuscamman}
 \email{eneuscamman@berkeley.edu}

  \affiliation{ 
Department of Chemistry, University of California, Berkeley, California 94720, USA%\\This line break forced with \textbackslash\textbackslash
}
\affiliation{%
Chemical Sciences Division, Lawrence Berkeley National Laboratory, Berkeley, California 94720, USA%\\This line break forced% with \\
}%

\date{\today}% It is always \today, today,

\begin{abstract}
We present an approach for augmenting Gaussian atomic orbitals with correct nuclear cusps.
Like the atomic orbital basis set itself, and unlike previous cusp corrections,
this approach is independent of the many-body method used to prepare wave functions
for quantum Monte Carlo.
Once the basis set and molecular geometry are specified, the cusp-corrected atomic
orbitals are uniquely specified,
regardless of which density functionals, quantum chemistry methods, or
subsequent variational Monte Carlo optimizations are employed.
We analyze the statistical improvement offered by these cusps in a number of molecules
and find them to offer similar advantages as molecular-orbital-based approaches
while maintaining independence from the choice of many-body method.
\end{abstract}

\maketitle

\section{\label{sec:level1}Introduction } 

%%%%%%%%%%%%%%%%%%%%%%%%%%%%%%%%%%%%%%%%%%%%%%%%%%%%%%%%%%%%%%%%%%%%%%%%%%%
% Intro paragraph about the importance of basis function choice.
%%%%%%%%%%%%%%%%%%%%%%%%%%%%%%%%%%%%%%%%%%%%%%%%%%%%%%%%%%%%%%%%%%%%%%%%%%%

The choice of functions to include in the one-electron basis set is
critical in electronic structure theory.
In general, a good basis set offers efficient numerical calculation,
appropriate flexibility for the physics at hand, and a form that
is straightforwardly transferrable between molecules.
In molecular electronic structure, examples of important flexibility
include the use of polarization functions to angularly resolve
molecular orbital (MO) features \cite{dunning1971} and the use of
diffuse functions to capture the physics of anions \cite{spitznagel1982}
and Rydberg states. \cite{kaufmann1989}
To achieve efficient calculation, quantum chemistry has long relied
on atom-centered Gaussian atomic orbital (AO) basis sets,
which greatly simplify the evaluation of key integrals. \cite{helgaker2013book}
Thanks to their dependence on only the identities and positions of
the nuclei, such Gaussian basis sets have proven readily transferrable
between different molecules.
However, the incorrect shape of Gaussian orbitals in the immediate vicinity
of the nucleus makes them less desirable in all-electron real-space
quantum Monte Carlo methods, whose accuracy and efficiency rely on
the hydrogenic divergence of kinetic energy (KE) to counteract the diverging
Coulomb energy at the nucleus. \cite{Foulkes}
Given the high accuracy offered by Monte Carlo methods, a way to
correct orbital shapes at nuclei that retains the efficiency,
flexibility, and transferability of the parent basis set
is highly desirable.
\blfootnote{${}^\dagger$\ These authors contributed equally to this work.}

%%%%%%%%%%%%%%%%%%%%%%%%%%%%%%%%%%%%%%%%%%%%%%%%%%%%%%%%%%%%%%%%%%%%%%%%%%%
% Overview of VMC and motivation based on its strengths
%%%%%%%%%%%%%%%%%%%%%%%%%%%%%%%%%%%%%%%%%%%%%%%%%%%%%%%%%%%%%%%%%%%%%%%%%%%

Variational Monte Carlo (VMC), for example, has proven capable of delivering
high accuracy in a number of contexts that challenge more traditional
approaches like density functional theory (DFT) and ground state
coupled cluster (CC).
These include charge transfer and doubly excited states,
\cite{LeonRev, Shepard}
core excitations, \cite{Scott2020}
thiophene, \cite{Dash2021}
and the carbon dimer. \cite{Ludovicy2023}
Success in these areas is largely due to VMC's ability
to combine Jastrow factors
with a configuration interaction expansion carefully
selected from quantum chemistry methods,
\cite{Clark2011, Assaraf2017, Petruzielo, Giner2016, Dash2019, Dash2021, Sergio2019, Leon2020}
although much recent work has also investigated
neural network wave function forms that have proven
particularly accurate in small molecules.
\cite{Ferminet2020, Paulinet2020, MLrev2023}
In both approaches, it can be useful and sometimes critical
to employ a quantum-chemistry-derived initial guess for
the molecular orbitals (MOs) and other wave function components,
and so the ability to interface effectively with Gaussian basis
sets is crucial.

%%%%%%%%%%%%%%%%%%%%%%%%%%%%%%%%%%%%%%%%%%%%%%%%%%%%%%%%%%%%%%%%%%%%%%%%%%%
% Paragraph discussing the standard approach: pseudopotentials and ECPs
%%%%%%%%%%%%%%%%%%%%%%%%%%%%%%%%%%%%%%%%%%%%%%%%%%%%%%%%%%%%%%%%%%%%%%%%%%%

The most widely used approach to make VMC compatible with Gaussian basis sets
and to deal with energy divergences near nuclei more generally
is to employ pseudopotentials, also known as effective core potentials (ECPs).
\cite{STU, BFD, cECP, ccECP}
In these approaches, the core electrons are removed and the divergent
Coulomb potential near the nucleus is replaced with an effective potential
designed to replicate the effect of the nucleus and core electrons on the
valence electrons.
As pseudopotentials lack Coulomb divergences, they remove the need for
the orbital basis to produce KE divergences, making VMC directly compatible
with Gaussian basis sets.
Furthermore, core electrons typically make the largest contributions to
VMC's energy variance, and so pseudopotentials also tend to reduce
statistical uncertainty.
They can also account for relativistic effects, \cite{dolg2012relativistic}
which can improve the energy compared to all-electron
non-relativistic calculations on heavier atoms. \cite{MLECP}
However, despite these clear advantages, pseudopotentials
require careful management of the localization error
\cite{Casula2006, Krogel, Scemama2019}
and can be less accurate than all-electron calculations.
For example, Wang et al.\ found that all-electron calculations for
first row atoms outperformed those using pseudopotentials when
evaluating ionization potentials. \cite{MLECP}
More directly, pseudopotentials cannot be used to study processes involving
core electrons, such as X-ray absorption spectroscopy. \cite{Scott2020}
Thus, while pseudopotentials are highly effective in many applications,
all-electron calculations remain preferred for some.
In those settings, electron-nuclear Coulomb divergences are present,
and Gaussian basis sets alone are no longer sufficient.

%%%%%%%%%%%%%%%%%%%%%%%%%%%%%%%%%%%%%%%%%%%%%%%%%%%%%%%%%%%%%%%%%%%%%%%%%%%
% Paragraph discussing cusp conditions and STO basis sets
%%%%%%%%%%%%%%%%%%%%%%%%%%%%%%%%%%%%%%%%%%%%%%%%%%%%%%%%%%%%%%%%%%%%%%%%%%%

To avoid energy divergences in all-electron calculations, the wave function
must satisfy cusp conditions that ensure the KE diverges in a way that
cancels the Coulomb divergence as particles coalesce.
\cite{katoEigenfunctionsManyparticleSystems1957a}
In practice, electron-electron cusps are typically satisfied using Jastrow factors.
\cite{Drummond2004}
Electron-nuclear cusps can also be satisfied with Jastrow factors, but
doing so often makes wave function optimization more difficult.
\cite{Drummond2004, Scott2020}
Alternatively, electron-nuclear cusps can be incorporated directly into
the orbitals.
One widely used approach is to replace Gaussian basis functions with a
Slater-type orbital (STO) basis, \cite{STO} which by construction provide
the necessary non-analytic features for cusps.
However, employing STOs makes communicating with Gaussian-based quantum
chemistry packages more difficult, and most do not support STOs directly
(although there are exceptions \cite{AMS2001}).
Furthermore, even if an individual STO satisfies a particular cusp
condition, linear combinations of STOs with different exponents,
necessary for constructing accurate molecular orbitals, may not,
which can further complicate wave function optimization within VMC.

%%%%%%%%%%%%%%%%%%%%%%%%%%%%%%%%%%%%%%%%%%%%%%%%%%%%%%%%%%%%%%%%%%%%%%%%%%%
% Paragraph about adding cusps to Gaussians
%%%%%%%%%%%%%%%%%%%%%%%%%%%%%%%%%%%%%%%%%%%%%%%%%%%%%%%%%%%%%%%%%%%%%%%%%%%

To maintain easy compatibility with Gaussian and plane wave basis sets
while achieving correct cusps at the nuclei, multiple groups have taken
the approach of adding cusps to MOs that initially lack them.
\cite{Ma2005, Per2008, Manten2001, Loos2019, nakatsuka}
These approaches significantly lower the energy variance in VMC,
and are directly compatible with Gaussian-type orbitals (GTOs).
However, the precise outcomes of these MO-based approaches,
along with similar methods that add cusps to the electron density,
\cite{Kussman2007}
depend on the shapes of the MOs and therefore on the many-electron method
(e.g.\ density functional) that was chosen that produced them.
This method dependence stands in sharp contrast to the situation for AO basis
sets, whose transferability stems in part from the fact that, once
a basis is chosen, its properties depend only on the molecular geometry.
In order to remove this method dependence while maintaining
compatibility with the GTOs of quantum chemistry,
we therefore explore in the present study the effectiveness of
adding cusps directly to the AO basis functions themselves.

%%%%%%%%%%%%%%%%%%%%%%%%%%%%%%%%%%%%%%%%%%%%%%%%%%%%%%%%%%%%%%%%%%%%%%%%%%%
% Paragraph describing goal of this paper and its layout
%%%%%%%%%%%%%%%%%%%%%%%%%%%%%%%%%%%%%%%%%%%%%%%%%%%%%%%%%%%%%%%%%%%%%%%%%%%

Specifically, we study an approach to constructing cusped Gaussian AOs
in which each AO is cusped at each nuclei.
In addition to making VMC's orbital basis set independent of the quantum
chemistry method that supplies the initial wave function, it also ensures
that the basis functions remain localized in space, a property that helps
avoid unnecessary orbital evaluation costs in larger systems.
Further, as with cusped MO approaches, any linear combination of these
basis functions will also satisfy the nuclear cusps, simplifying the
task of orbital optimization within VMC.
To begin, we will describe our theoretical approach and its instantiation
in a lightweight open-source software library, Cusping Gaussian Atomic Orbitals with Slaters (CGAOWS).
We then test the cusped AOs in a collection of small molecules and analyze
the results, finding that the good performance of previous MO-based methods
can be maintained while simultaneously freeing the VMC orbital basis
from any dependence on the many-electron method used to produce the
initial MOs.

\section{Theory}

For a wave function to obey Kato's cusp condition
\cite{katoEigenfunctionsManyparticleSystems1957a}
as an electron approaches nucleus $I$, it must satisfy
\begin{equation} \label{ddr_psi_eq_Z_psi}
\begin{split}
    \left. \frac{\partial \langle \Psi \rangle}{\partial r} \right|_{r=0}
      = -Z_I \, \langle \Psi \rangle_{r=0}
\end{split}
\end{equation}
\noindent where $\langle \Psi \rangle$ is the average value of the wave function on the
surface of the radius-$r$ sphere centered on nucleus $I$.

If we wish the cusp condition in Eq.~\ref{ddr_psi_eq_Z_psi} be satisfied
by the Slater determinant, then one may equivalently express the
cusp condition in terms of the AOs $\chi_i$ within the linear combination of
atomic orbitals (LCAO) construction of each MO,
$\phi_a(r) = \sum_i c_{ia} \chi_i(r)$.
\begin{equation} \label{ddr_chi_eq_Z_chi}
\begin{split}
    \left. \frac{\partial \langle \chi_i \rangle }{\partial r}\right|_{r=0}
      = -Z_I \, \langle \chi_i \rangle_{r=0} \quad \forall \hspace{2mm} i,I
\end{split}
\end{equation}
If this equation is satisfied for all AO-nucleus pairs, including in the tails of
AOs centered on other nuclei, then the Slater determinant of MOs will satisfy
the original cusp condition.

\begin{figure}[ht]
    \centering
    \includegraphics[width=0.9\linewidth]{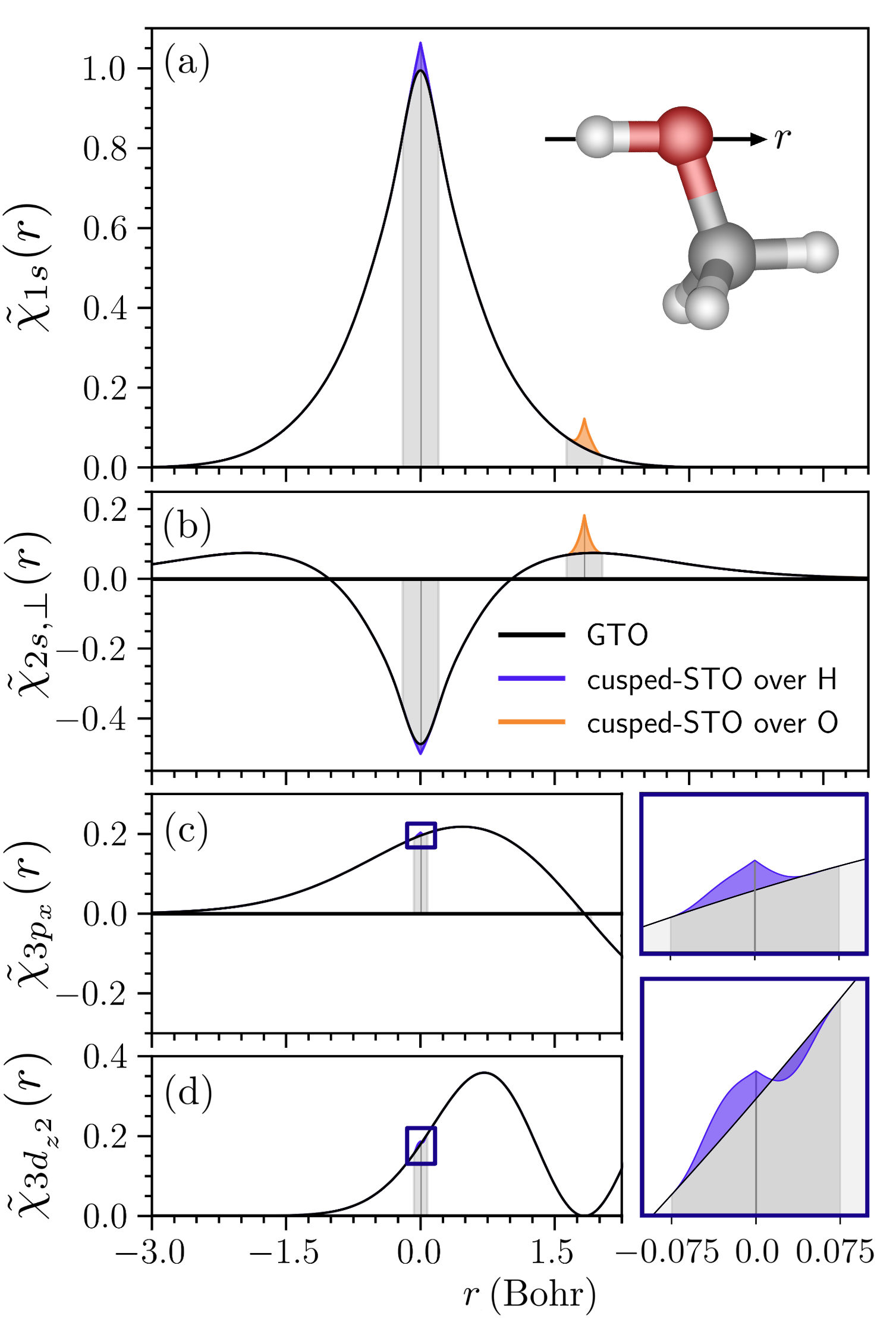}
    \caption{
    Visualization of the cusp corrected orbitals ($\tilde{\chi}(r)$) from a 6-31G(\textit{d}) basis evaluated through the H-O bond in methanol, where H is the origin, see top right molecule. Cusp correction over H (O) correspond to purple (orange). (a) H $1s$ orbital, (b) H $2s$ orbital orthogonalized against the 1s, (c) O $3p_{x}$ orbital, and (d) O $3d_{z^2}$. A zoomed in view of the cusps on the tails of (c,d) are to the right of their respective plots.
    }
    \label{fig:cusp}
\end{figure}

To address GTOs' failure to satisfy Eq.~\ref{ddr_chi_eq_Z_chi},
we introduce a modification that smoothly transitions from the original GTO to
a cusp-satisfying Slater-type function (cusped-STO) within a cusp radius
$r_{c}$ for each AO at each nuclear center.
Any linear combination of cusp-satisfying AOs will thus
result in a similarly cusp-corrected MO -- as long as each AO
is corrected over all nuclei in which it has
a non-negligible magnitude.
This approach (Fig.~\ref{fig:cusp})
(i) is MO-independent
(ii) is independent of the method used to construct the initial trial wave function
(iii) accommodates GTO basis sets from quantum chemistry, and
(iv) can be applied before starting the QMC calculation.

Within a radius $r_c$ sphere around each nucleus,
we replace each GTO basis function $\chi_n$
with a cusp-corrected basis function $\tilde{\chi}_n$ that is defined
within the $r_c$ sphere as 
\begin{equation}\label{simplified_cusp_GTO}
 \tilde{\chi}_n(r) = (1-b(r)) \, \chi_n(r) + b(r) \, Q(r; \vec{q}_{nI})
\end{equation}
\noindent where $r$ is the distance to the nuclei in question and $b(r)$ is the $5$th-order polynomial
\begin{equation} \label{b5}
\begin{split}
   b(r) &= c_1 r^5 + c_2 r^4 + c_3 r^3 + c_4 r^2 + c_5 r + c_6 
\end{split}
\end{equation}
that gradually switches from the GTO at $r=r_c$ to the cusped-STO,
$Q(r; \vec{q})$, at $r=0$.
Outside the cusp radius, the original GTO is left unchanged.
For simplicity, and because we anyways want to disturb the original GTO as little
as possible to maintain the character of the original basis set, we constrain the
cusp radii so that the correction regions around different nuclei do not overlap.
The precise coefficients within the switching function $b(r)$ are chosen as
$c_1 = -{6}/{r_c^5}$, $c_2 = {15}/{r_c^4}$, $c_3 = -{10}/{r_c^3}$,
$c_4 = 0$, $c_5 = 0$, and $c_6 = 1$,
in order to ensure that $b(0)=1$, that $b(r_c)=0$,
and that $b'(0) = b''(0) = b'(r_c) = b''(r_c) = 0$.
This choice makes the updated orbital's first and second derivatives
continuous at $r=r_c$ and preserves the cusped-STO's satisfaction of
the cusp condition at $r=0$.
It also ensures that, at the $r=0$ and $r=r_c$ boundaries, the KE contribution
is coming entirely from the STO or GTO, respectively.

In Eq.~\ref{simplified_cusp_GTO}, every AO is cusp-corrected around each nucleus
at which it has an appreciable magnitude,
meaning a value whose magnitude is at least $10^{-15}$ times greater than the orbital's maximum magnitude.
Thus, energy divergences are handled for AOs at their own nucleus and
at other nuclei, as shown in Figs.~\ref{fig:cusp}c and \ref{fig:cusp}d
(note that the magnitude cutoff means that we do not add
cusps to $p$ or $d$ orbitals at their own nuclei).
An advantage to smoothly switching between the GTO and a cusped-STO in the
cusping region via a linear combination of the two is that the spherical asymmetry
about a nucleus in the tail of a GTO (Fig.\ref{fig:cusp}c,d) can be maintained
within $\tilde{\chi}(r)$ such that the original GTO shape is not too strongly altered.
Meanwhile, the spherical symmetry of the cusped-STO dominates near the nucleus
and satisfies the necessary cusp condition.

To construct the cusped-STO function $Q(r)$ for a specific GTO-nucleus pair,
we begin with the overall goal
that it satisfy the cusp condition so that the KE will
cancel the electron-nuclear Coulomb divergence at the nucleus in question.
\begin{equation}\label{KE_eq_eN}
\begin{split}
\begin{aligned}
    -\frac{1}{2} \frac{\nabla^2 Q(r)}{Q(r)} &= \frac{Z}{r}
\end{aligned}
\end{split}
\end{equation}
A natural starting point is the hydrogenic 1s orbital for that nucleus,
and so we define an initial form $Q(r;~q_0)$ for $Q$ as
\begin{equation}\label{cusp_Q_slater}
\begin{split}
\begin{aligned}
    Q(r;~{q}_0) &= q_{0} \, e^{-Z r} 
   \end{aligned}
\end{split}
\end{equation}
in which the constant $q_0$ controls how much the cusp correction feature
sticks out from the original GTO (see Fig.~\ref{fig:cusp}).
We determine $q_0$ for each GTO-nucleus pair by minimizing the
one-electron energy functional
\begin{equation} \label{1_elec_energy_per_nuc}
\begin{split}
  \tilde{E} &=
  \frac{   \bra{\chi_n}\hat{H}\ket{\chi_n}
         - \bra{\chi_n}\hat{H}\ket{\chi_n}_{r<r_c}
         + \bra{\tilde{\chi}_n}\hat{H}\ket{\tilde{\chi}_n}_{r<r_c}
       }{
           \braket{\chi_n}{\chi_n}
         - \braket{\chi_n}{\chi_n}_{r<r_c}
         + \braket{\tilde{\chi}_n}{\tilde{\chi}_n}_{r<r_c}
        } 
\end{split}
\end{equation}
\noindent with the Broyden–Fletcher–Goldfarb–Shanno algorithm.
\cite{Broyden,Fletcher,Goldfarb,Shanno}
In Eq.~\ref{1_elec_energy_per_nuc}, the denominator accounts for the AO's normalization,
$\hat{H}$ is the one-body Hamiltonian
\begin{equation}\label{H}
 \hat{H} = -\frac{1}{2}\nabla^2_e - \sum_I \frac{Z_I}{|\vec{r}_e - \vec{R}_I|}
\end{equation}
for the electron $e$, % and the electron-nuclear potential contains all nuclei in the system. 
and the $r<r_{c}$ subscript indicates that the integrals within the indicated
expectation values are truncated to the region of space within
the radius $r_c$ sphere about the nucleus in question.
Essentially, these one-electron optimizations choose the $q_0$
value for each GTO-nucleus pair that minimizes a simple approximation
of the AO energy.

\begin{figure}[htbp]
    \centering
    \includegraphics[width=0.9\linewidth]{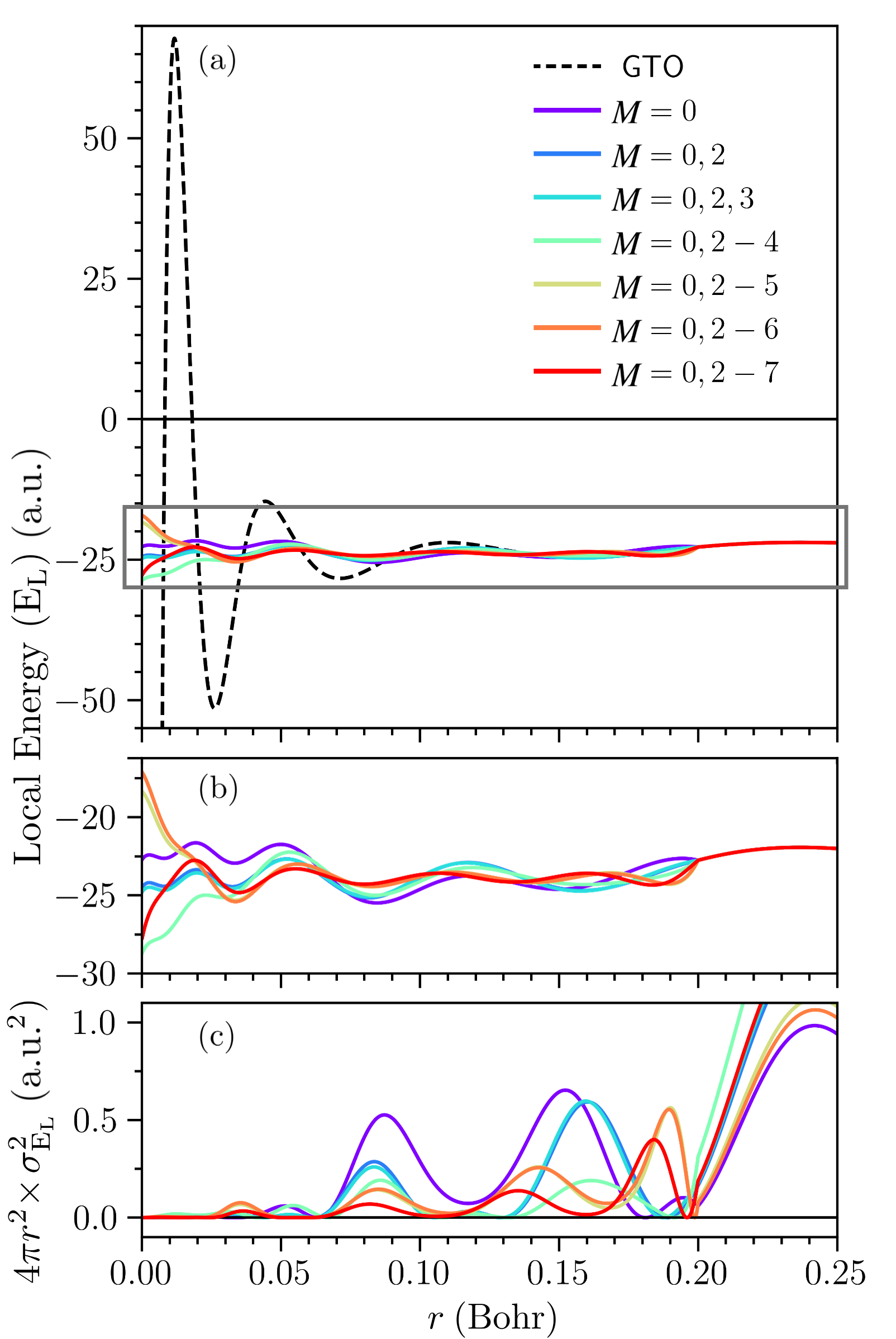}
    \caption{
    Local energy (a.u.) (a) of the localized core carbon MO in CH\textsubscript{3}OH, 6-31G(\textit{d}) basis, as an electron nears the carbon nucleus ($r\rightarrow0$) of the non-cusp corrected GTO (black dashed) and the cusp corrected orbitals (solid colors) as a function of the indices included in the Slater-like basis expansion of Eq.~\ref{cusp_Q_lc} ($n$). $r_c = 0.20$ Bohr.
    (b) Magnified plot of the cusp corrected AOs E\textsubscript{L} (a.u.) as a function of $n$. 
    (c) The contribution of the cusped orbital to the variance of the local energy ($\sigma^2_\text{E\textsubscript{L}}$ a.u.$^2$) in the sphere around the carbon. 
    }
    \label{fig:El_v_r}
\end{figure}

While this initial choice of $Q=Q(r;~q_0)$ satisfies the cusp-condition at $r=0$,
we find that it does a poor job of managing KE fluctuations in the
intermediate region in between $r=0$ and $r=r_c$.
We therefore generalize this initial choice by including expanded
radial flexibility in our final form for $Q$.
\begin{equation}\label{cusp_Q_lc}
\begin{split}
\begin{aligned}
    Q(r; \vec{q}) &= q_0 \, e^{-Z r} + \sum_{i=2}^M q_{i} \, {r}^i \, e^{-Z r} 
   \end{aligned}
\end{split}
\end{equation}
Note that, while we include terms up to $M=7$, we omit the term linear in $r$,
as its derivatives would interfere with satisfying the cusp condition. 
We determine the coefficients $\vec{q}$ through variational optimization
($\mathbf{H}\,\vec{q} = E\,\mathbf{S}\,\vec{q}$)
within the radius $r_c$ sphere around the nucleus.
In this optimization, we simplify even further, neglecting the space outside $r_c$
entirely and including only the kinetic energy and the attraction to
the nucleus in question.
The Hamiltonian and overlap matrices $\mathbf{H}$ and $\mathbf{S}$
are constructed in the basis
$\{(1-b(r))\chi_n(r),\,\, b(r)\,e^{-Z r}r^i;~ i=0,2,3,\ldots\}$,
and the resulting lowest-energy eigenvector $\vec{q}$ is scaled so that
the coefficient on the $(1-b(r))\chi_n(r)$ term of Eq.~\ref{simplified_cusp_GTO} equals one,
which ensures that the boundary conditions at $r_c$ remain satisfied. 

In Fig.~\ref{fig:El_v_r}, we show how the local energy E\textsubscript{L}
of the localized $1s$ carbon MO in methanol behaves
for different truncations in the order of $M$.
A true energy eigenstate would have a constant E\textsubscript{L},
and therefore zero-variance. \cite{assarafZeroVariancePrincipleMonte1999}
The cusped orbital should ideally make E\textsubscript{L} finite everywhere and
minimize the amplitudes of the ``wiggles'' seen in the GTO's E\textsubscript{L}
in order to both satisfy the central limit theorem and reduce
the variance within a VMC calculation.
Note that, for $M \geq 2$, there are larger fluctuations of E\textsubscript{L}
near the origin but smaller fluctuations in the remainder of the $r<r_c$ region.
Since the probability of sampling very close to the nucleus (e.g.\ $r<0.02$)
is much smaller than sampling in the remainder of the $r<r_c$ region,
this behavior is an expected trade off made by the variational
optimization that should reduce the overall VMC energy variance.
Indeed, by plotting the net contributions to the variance at different radii
(Fig.~\ref{fig:El_v_r}c), we see that this trade off is exactly what is happening,
although we also note that the ability to reduce variance contributions near
$r=r_c$ ($r_c=0.2$ in this example) is limited by the constraint
that we match the GTO at the boundary, limiting the final orbital quality
there to that of the original GTO.

The expansion with $M=7$ (red, Fig.~\ref{fig:El_v_r}a,b) was chosen for the final
form of $Q(r)$, as it most significantly decreases the cusp region's contribution
to the energy variance $\sigma^2$, as demonstrated in Fig.~\ref{fig:El_v_r}c. 
Beyond $M=7$, the basis functions rapidly become linearly dependent,
limiting the quality of the optimization.
In principle, one could adopt a different basis in which the functions are orthogonal
by construction (e.g.\ Chebyshev polynomials), but, as we will see in the general results,
stopping at $M=7$ is sufficient to produce results comparable to previous MO-based
cusping schemes.  

In our cusping package, \textsc{CGAOWS},
we have chosen specific cusp radii for all nuclei up to $Z=10$ so that the cusp
corrections do not overlap and do not disturb the original GTO shapes more than
necessary, so as to retain as closely as possible the character of the
original basis.
Our choices were guided by using methanol as a test case, in which we sought
to find a balance between variance and average E\textsubscript{L} reduction
while maintaining generalizability across the various AO and $Z$ combinations.
The default cusp radius for $s$ orbitals is $r_c=0.2$ Bohr, except
in the case of a hydrogen $s$ orbital being cusped at a hydrogen
nucleus, in which case $r_c=0.1$ Bohr.
For $p$ or $d$ orbitals, the default is $r_c=0.075$ Bohr.
These parameters were chosen given that they produce a satisfactory reduction in the variance, however, further optimization is possible. They can be readily tuned within the
software for specific molecular systems and basis sets if desired.

%\subsubsection{Orbital orthogonalization}
Finally, in order to increase the efficacy of the chosen cusp radii,
we redefine $s$ orbital GTOs with principle quantum number 2 or higher
by Gram-Schmidt orthogonalizing them against their corresponding $1s$ orbital.
Note that this does not change the span of the basis set, but it does produce
more uniformity in the shapes of GTO $s$ orbitals near their centers,
making a one-size-fits-all cusp radius for $s$ orbitals more effective.
An example of this approach is seen in Fig.~\ref{fig:cusp}b, where
the $r=0$ behavior of the $2s$ basis function has been made similar to that of the $1s$
function via orthogonalization.
Note that, as seen in Figs.\ \ref{fig:cusp}c and \ref{fig:cusp}d,
we do not orthogonalize against other nuclei's $1s$ orbitals when adding cusps
for those nuclei, as this would decrease the locality of the basis functions.
Happily, we find that the default cusp radius, along with the optimization
of $Q(r)$, produces good results for those cusps.
For an $s$ orbital's own nucleus, however, we perform the orthogonalization
as we find that it significantly improves the quality of the results.
Of course, when importing initial MOs from quantum chemistry into VMC, the
LCAO coefficient matrix must be updated to reflect this redefinition of
higher principle quantum number $s$ basis functions.

\section{Results}
%
% Results roadmap
%
The performance of the presented cusping scheme is tested using a restricted single Slater determinant trial wave function constructed with Foster-Boys\cite{fosterCanonicalConfigurationalInteraction1960} localized Hartree-Fock orbitals from \textsc{Pyscf}\cite{sunLibcintEfficientGeneral2015,sunPySCFPythonbasedSimulations2018,sunRecentDevelopmentsPySCF2020} and a 6-31G($d$) Pople basis set. \cite{Pople} % pyscf citation checked and g2g
By using a single Slater wave function without a Jastrow factor, all benefit in variance reduction may be attributed to the cusp-correction.
All VMC calculations from this work were performed on molecular geometries optimized at the MP2/6-31G(\textit{d}) level of theory.\cite{cccdbd}
As tests cases, neon (Fig.~\ref{fig:El_thru_Ne}) and methanol (Fig.~\ref{fig:El_thru_H3COH}) are used to demonstrate the impact of this cusp correction scheme on atomic and molecular systems, respectively, by evaluating E\textsubscript{L} as an electron ``walks'' across the nucleus.
Following this, results on a test set of 16 small gas phase molecules are presented in Tables~\ref{table:normal_dat} and~\ref{table:nonnormal_dat}; the former containing statistical analysis assuming normally-distributed data (\textit{i.e.\ }mean energy and variance) and the latter assuming non-normally distributed data (\textit{i.e.\ }median energy, interquartile range, and range). 
Four of the molecules in this test set including LiH, C\textsubscript{2}H\textsubscript{6}, N\textsubscript{2}, and CO\textsubscript{2}, chosen based on data availability, are compared against other relevant cusping schemes in Table~\ref{table:normal_dat}.
All VMC calculations from this work were computed
using our own VMC software, augmented by the newly developed \textsc{CGAOWS} package.
For details about the software, see the SI.

\subsection{Neon and Methanol}\label{sec:Ne}
\begin{figure}[t]
    \centering
    \includegraphics[width=0.9\linewidth]{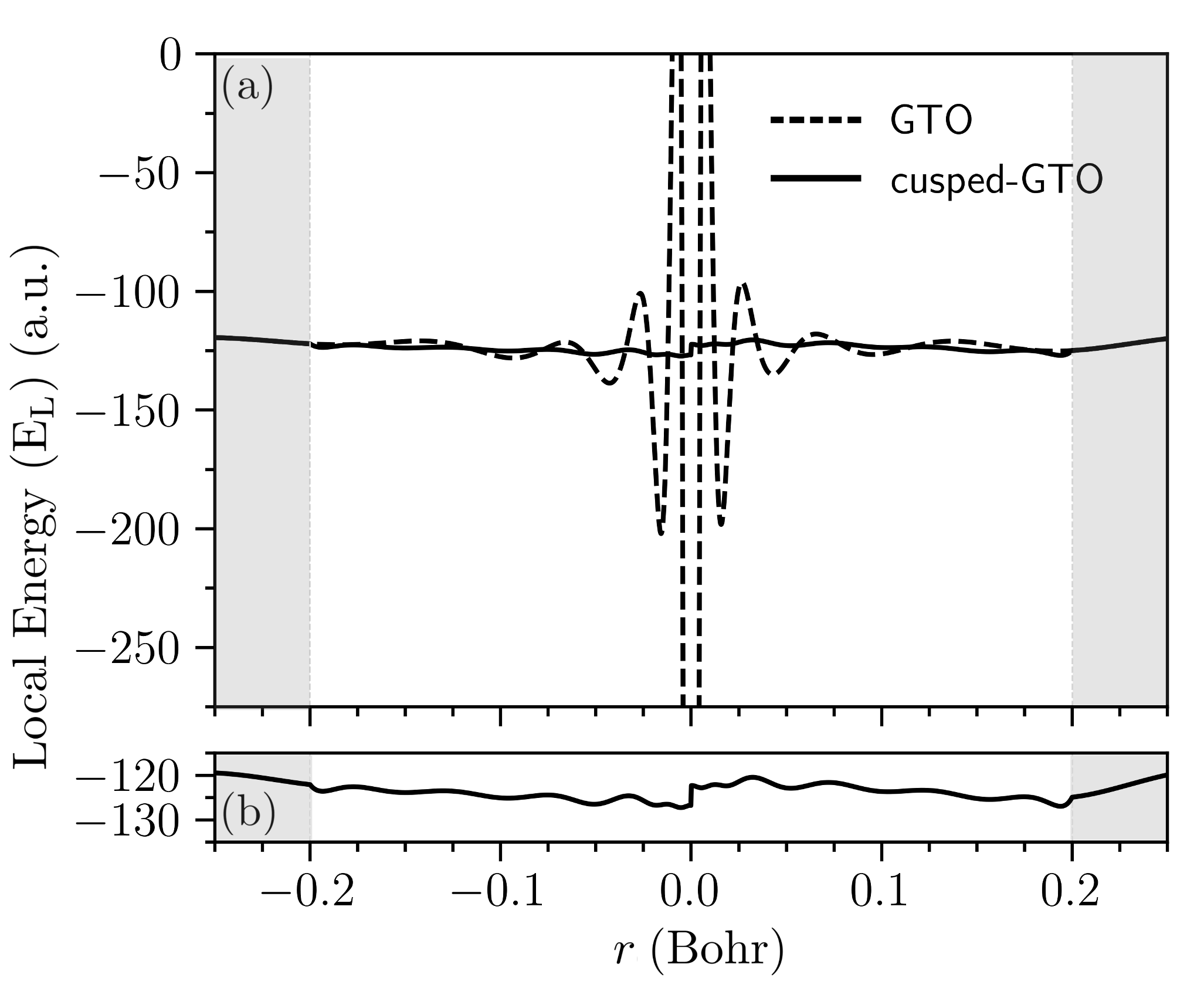}
    \caption{
    Local energy (a.u.) of single Slater determinant wave function, no Jastrow factor,  as an electron ``walks'' across the nucleus of a neon atom, in a 6-31G($\textit{d}$) basis. $r_c=0.2$ Bohr for all AOs, the shaded area outside of $|r|>r_c$ indicates region of no cusp-correction. (a) Uncusped GTO basis (black dashed) and cusp-corrected basis (black solid). (b) Magnified E\textsubscript{L} of the cusp-corrected basis.
    %\textcolor{blue}{Add subplots with different orbital contribution breakdown}
    }\label{fig:El_thru_Ne}
\end{figure}
As an initial demonstration of the presented cusping scheme, Fig.~\ref{fig:El_thru_Ne} depicts E\textsubscript{L} of the Ne atom as an electron is moved through the nucleus, holding all other electrons fixed. 
An initial VMC calculation, with appropriate burn-in, was performed to produce an electron configuration distributed with respect to the wave functions probability distribution. The alpha spin electron closest to the nucleus was then repositioned to move through the nucleus along the $x$-axis. 
Fig.~\ref{fig:El_thru_Ne} compares both the uncusped- (dashed line) and cusped- (solid line) GTO bases.
While the details of the fluctuations in E\textsubscript{L} for both cases depend on the positions of the other electrons, the general behavior is consistent when other configurations are put through the same test.
Note, outside of neon's cusp radius of $0.2$ Bohr the cusped-GTO and GTO appear to return to evaluating the same orbital, however, in actuality there is a slight, unobservable difference between the two.
This difference is expected, as it is caused by other electrons
residing within the cusp radii of the atom in this particular configuration of the electrons, as would be expected in atomic or molecular systems.

As expected, the GTO basis yields significant oscillations and a divergence of E\textsubscript{L} as the electron nears the nucleus. In the cusped-GTO basis, these oscillations are dramatically reduced and E\textsubscript{L} remains finite everywhere. This result closely resembles the behavior of the MO-based cusping scheme shown in Fig.~4 of Ma et al.\cite{Ma2005} 
The remaining subtle ``wiggles'' visible in the cusped-GTO case (Fig.~\ref{fig:El_thru_Ne}b) are a byproduct of the choice of polynomial switching function in Eq.~\ref{b5} that controls the interpolation between the pure GTO and the cusped-STO of Eq.~\ref{cusp_Q_lc}.
Two paths could be taken to further reduce the remaining
fluctuations in E\textsubscript{L} near the nucleus.
First, a higher-order polynomial expansion, using an orthogonal construction
to avoid linear dependency issues, could be employed.
Second, adding similar corrections to non-$s$ basis functions
in the $r<r_c$ region, although not necessary to cancel any divergences,
would improve their kinetic energy behavior near the nucleus.
We forgo these steps in the present study, because the $M=7$ expansion
applied only to $s$ orbitals already achieves reductions
comparable to previous MO-based schemes. \cite{Ma2005}

\begin{figure}[t]
    \centering
    \includegraphics[width=0.9\linewidth]{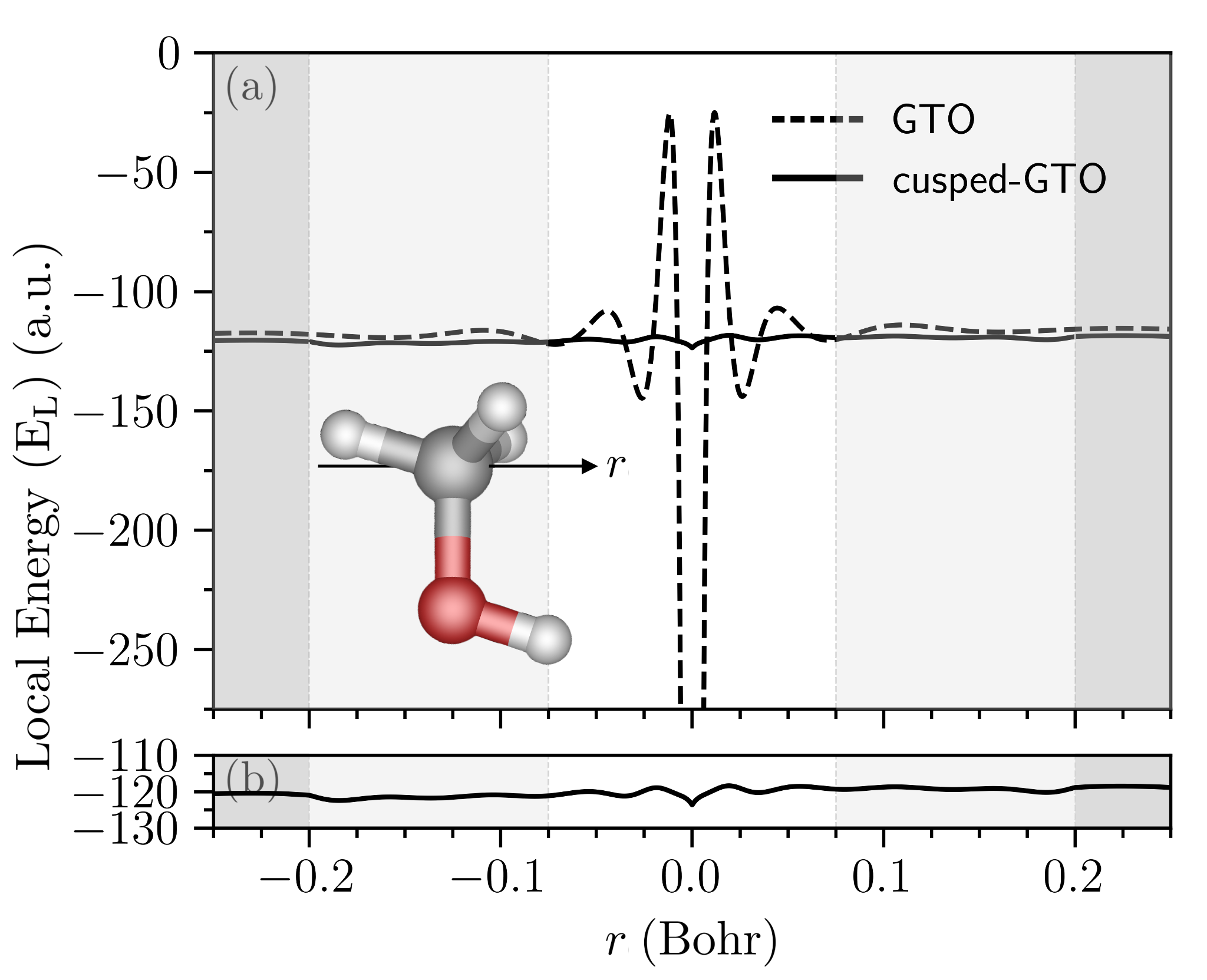}
    \caption{
    Local energy (a.u.) of single Slater determinant wave function, no Jastrow factor, as an electron  ``walks'' across the carbon nucleus of methanol, see slice through molecule for the trajectory, in a 6-31G($\textit{d}$) basis. 
    Light gray shading outside of $|r|=0.075$ indicates the cusp radii around the carbon nucleus for the oxygen atom's $p$ and $d$ orbitals. The dark gray shading outside of $|r|=0.2$ indicates the cusp radii
    around the carbon nucleus for all other basis functions. 
    }\label{fig:El_thru_H3COH}
\end{figure}

To verify that similar behavior is achieved in a molecular system,
where adding cusps for other nuclei in the tails of each AO now
matters, we perform a similar test in methanol.
Using the same procedure as in neon, Fig.~\ref{fig:El_thru_H3COH} depicts E\textsubscript{L} as an electron moves through the carbon nucleus.
As in neon, we see a large reduction in local energy fluctuations
compared to the original GTO basis.
More significantly than neon, we see for methanol a difference in the
local energy values of the cusped and uncusped results outside
of the carbon atom's outermost cusp radius of $0.2$ Bohr as well.
Indeed, as one looks at larger and larger molecules, the chances
of sampling a configuration in which none of the electrons lie
within any of the nuclei's cusp radii becomes vanishingly small.

\subsection{Standard test set of small molecules}
\begin{table*}[htbp] 
\centering
\caption{Comparison of HF energies ($E_\text{HF}$), VMC energies ($E_\text{VMC}$) with and without electron-nuclear cusps, variances ($\sigma^2$), and standard errors ($\pm$ following VMC results), all in a.u., for this work and work by Kussman \textit{et al.}\cite{Kussman2007}, Ma \textit{et al.}\cite{Ma2005}, and Per \textit{et al.}\cite{Per2008} All results employed a Hartree-Fock-based single Slater determinant with no Jastrow factor. 
See text for geometry, basis set, and sampling details.}\label{table:normal_dat}
{\renewcommand{\arraystretch}{1.1} % for the vertical padding
\begin{tabular}{@{\extracolsep{7pt}}lcrrrrrc@{}}
\hline
\hline\\ [-1em]
         &    &     & \multicolumn{2}{c}{Not cusp corrected} & \multicolumn{2}{c}{Cusp corrected} & \multicolumn{1}{c}{} \\[0.8ex]
\cline{4-5}\cline{6-7}\\ [-0.8em]
Molecule &  & \multicolumn{1}{c}{$E_\text{HF}$} & \multicolumn{1}{c}{Mean $E_\text{VMC}$} & \multicolumn{1}{c}{$\sigma^2$} & \multicolumn{1}{c}{Mean $E_\text{VMC}$} & \multicolumn{1}{c}{$\sigma^2$} & \multicolumn{1}{c}{Variance reduction (\%)}\\ [0.2em]
\hline\\ [-0.5em]
LiH  & This work & -7.9808 & -7.982 $\pm$ 0.001 & 8.3 $\pm$ 0.8 & -7.9817 $\pm$ 0.0006 & 1.68 $\pm$ 0.02 & 79.81 \\
 & Kussman & -7.9809 & -7.980 $\pm$ 0.001 & 7.14 $\pm$ 0.02 & -7.9826 $\pm$ 0.0006 & 1.76 $\pm$ 0.01 & 75.35 \\
 & Ma & -7.9859 & -7.984 $\pm$ 0.002 & 6.7 $\pm$ 0.9 & -7.9864 $\pm$ 0.0005 & 1.64 $\pm$ 0.03 & 75.52 \\
 & Per & -7.9855 & -7.9845 $\pm$ 0.0015 & 8 $\pm$ 2 & -7.98536 $\pm$ 0.00049 & 1.607 $\pm$ 0.016 & 79.91 \\ [1ex]
C\textsubscript{2}H\textsubscript{6}    & This work & -79.2280 & -79.24 $\pm$ 0.02 & 275 $\pm$ 50 & -79.231 $\pm$ 0.002 & 19.9 $\pm$ 0.3 & 92.78 \\
 & Kussman & -79.2285 & -79.22 $\pm$ 0.01 & 214 $\pm$ 3 & -79.237 $\pm$ 0.005 & 16.1 $\pm$ 0.4 & 92.48 \\
 & Ma & -79.2567 & -79.24 $\pm$ 0.02 & 156 $\pm$ 31 & -79.259 $\pm$ 0.002 & 18.0 $\pm$ 0.2 & 88.46 \\
 & Per &  \multicolumn{1}{c}{-} &  \multicolumn{1}{c}{-} &  \multicolumn{1}{c}{-} &  \multicolumn{1}{c}{-} &  \multicolumn{1}{c}{-} &  \multicolumn{1}{c}{-} \\ [1ex]
N\textsubscript{2}                      & This work & -108.9343 & -108.95 $\pm$ 0.03 & 453 $\pm$ 90 & -108.936 $\pm$ 0.004 & 28.4 $\pm$ 0.2 & 93.72 \\
 & Kussman & -108.9354 & -108.9 $\pm$ 0.1 & 390 $\pm$ 8 & -108.927 $\pm$ 0.006 & 25 $\pm$ 1 & 93.59 \\
 & Ma & -108.9710 & -108.96 $\pm$ 0.02 & 308 $\pm$ 76 & -108.973 $\pm$ 0.003 & 25.1 $\pm$ 0.3 & 91.85 \\
 & Per & -108.961 & -108.912 $\pm$ 0.030 & 276 $\pm$ 76 & -108.9677 $\pm$ 0.0033 & 24.76 $\pm$ 0.46 & 91.03 \\  [1ex]
CO\textsubscript{2}                      & This work & -187.6275 & -187.62 $\pm$ 0.04 & 774 $\pm$ 231 & -187.614 $\pm$ 0.005 & 53.0 $\pm$ 0.6 & 93.16 \\
 & Kussman & -187.6284 & -187.60 $\pm$ 0.01 & 598 $\pm$ 9 & -187.635 $\pm$ 0.009 & 45 $\pm$ 2 & 92.47 \\ 
 & Ma & -187.6880 & -187.62 $\pm$ 0.02 & 348 $\pm$ 39 & -187.691 $\pm$ 0.003 & 44.9 $\pm$ 0.5 & 87.10 \\  
 & Per &  \multicolumn{1}{c}{-} &  \multicolumn{1}{c}{-} &  \multicolumn{1}{c}{-} &  \multicolumn{1}{c}{-} &  \multicolumn{1}{c}{-} &  \multicolumn{1}{c}{-} \\ [0.5em]
 \hline\\ [-1em]
 & This work &   &   &   &   &    \\ [0.2em]
 \hline\\ [-0.5em]
%  Molecule                            E_HF            E_vmc              sig                   E_vmc                  sig           
Li\textsubscript{2}                 & \multicolumn{1}{c}{-} & -14.8664  & -14.869 $\pm$ 0.003  & 21 $\pm$ 4       &-14.8676 $\pm$ 0.0009  &3.18 $\pm$ 0.05  & 84.79  \\
CH\textsubscript{4}                 & \multicolumn{1}{c}{-} & -40.1947  & -40.182 $\pm$ 0.008  & 82 $\pm$ 13      &-40.1984 $\pm$ 0.0009  &9.9 $\pm$ 0.1    & 87.85  \\
C\textsubscript{2}H\textsubscript{2}& \multicolumn{1}{c}{-} & -76.8149  & -76.80 $\pm$ 0.01    & 198 $\pm$ 27     &-76.820 $\pm$ 0.002    &19.4 $\pm$ 0.7   & 90.20  \\
C\textsubscript{2}H\textsubscript{4}& \multicolumn{1}{c}{-} & -78.0306  & -78.03 $\pm$ 0.02    & 227 $\pm$ 44     &-78.030 $\pm$ 0.002    &19.1 $\pm$ 0.2   & 91.60  \\
HCN                                 & \multicolumn{1}{c}{-} & -92.8693  & -92.89 $\pm$ 0.01    & 398 $\pm$ 63     &-92.870 $\pm$ 0.003    &25 $\pm$ 1       & 93.82  \\
NH\textsubscript{3}                 & \multicolumn{1}{c}{-} & -56.1832  &  -56.17 $\pm$ 0.01   & 145 $\pm$ 19     &-56.189 $\pm$ 0.003    &15.5 $\pm$ 0.7   & 89.30  \\
H\textsubscript{2}CO                & \multicolumn{1}{c}{-} & -113.8629 & -113.78 $\pm$ 0.02   & 248 $\pm$ 24     &-113.858 $\pm$ 0.004   &31.0 $\pm$ 0.4   & 87.51  \\
CH\textsubscript{3}OH               & \multicolumn{1}{c}{-} & -115.0331 & -115.06 $\pm$ 0.03   & 646 $\pm$ 147    &-115.038 $\pm$ 0.003   &31.4 $\pm$ 0.7   & 95.14  \\
H\textsubscript{2}O                 & \multicolumn{1}{c}{-} & -76.0084  &  -76.01 $\pm$ 0.02   & 204 $\pm$ 24     &-76.006 $\pm$ 0.004    &22.3 $\pm$ 0.5   & 89.09  \\
H\textsubscript{2}O\textsubscript{2}& \multicolumn{1}{c}{-} & -150.7577 &  -150.74 $\pm$ 0.06  & 581 $\pm$ 133    &-150.744 $\pm$ 0.004   &43.8 $\pm$ 0.5   & 92.46  \\
O\textsubscript{2}                  & \multicolumn{1}{c}{-} & -149.5187 &  -149.7 $\pm$ 0.2    & 8732 $\pm$ 8148  &-149.512 $\pm$ 0.004   &43.6 $\pm$ 0.8   & 99.50  \\
HF                                  & \multicolumn{1}{c}{-} & -100.0002 & -99.97 $\pm$ 0.04    & 532 $\pm$ 200    & -99.992 $\pm$ 0.003   &32.5 $\pm$ 0.2   & 93.89  \\ [0.5em]
\hline \hline
\end{tabular}}
\end{table*}

\begin{figure}[ht]
    \centering
    \includegraphics[width=0.9\linewidth]{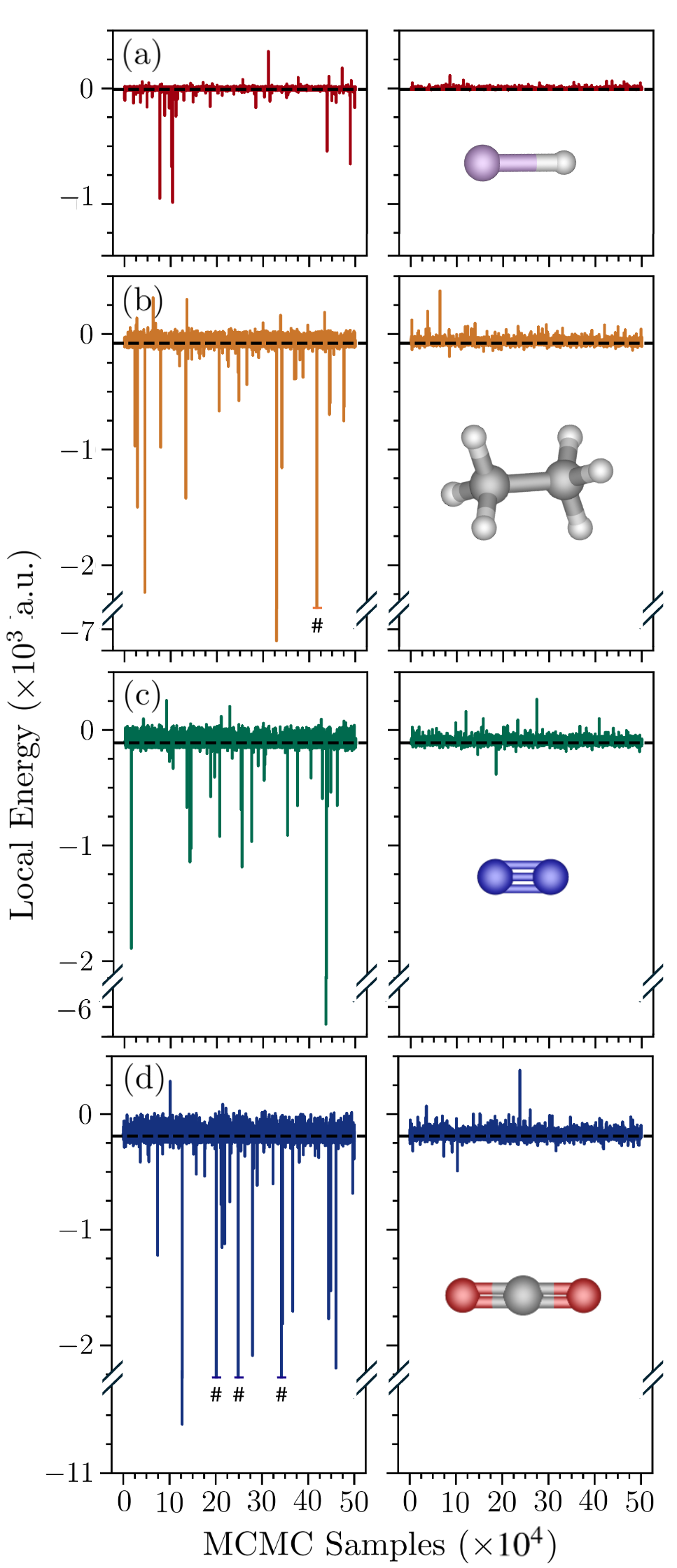}
    \caption{
    Local energies for 500,000 sample excerpts from (a) LiH (b) C\textsubscript{2}H\textsubscript{6} (c) N\textsubscript{2} (d) CO\textsubscript{2} (6-31G(\textit{d})). Left plot without cusps, right with cusps. As there is no Jastrow factor, the remaining outliers after nuclear cusp corrections are applied are explained by the lack of electron-electron cusps. 
    }
    \label{fig:El_v_MCMC}
\end{figure}
\begin{table}[htb]
\centering
\caption{Comparison of the median, interquartile range (IQR), and total range of the local energy in a.u.\ in uncusped and cusped VMC calculations for the 16 molecules studied in this work. The sample size in each calculation is 5 million.
} \label{table:nonnormal_dat} 
\resizebox{\columnwidth}{!}{%
    \begin{tabular}{@{\extracolsep{3pt}}lccrccrc@{}}
        \hline \hline\\[-0.8em]
        & \multicolumn{3}{c}{Not cusp corrected} & \multicolumn{3}{c}{Cusp corrected} & \\[0.8ex]
        \cline{2-4} \cline{5-7}\\ [-0.8em]
        Molecule &  Median & IQR & Range &  Median & IQR  & Range \\[0.8ex]
        \hline \\ [-1.ex]
       %  Molecule                          Median      IQR       Range      Median       IQR    Range 
        LiH                                 & -8.159   & 0.984  & 1509  & -8.192    & 0.923  & 256  \\[0.8ex]
        C\textsubscript{2}H\textsubscript{6}& -79.663  & 4.997  & 7653  & -79.913   & 3.982  & 1290 \\[0.8ex]
        N\textsubscript{2}                  & -109.410 & 6.197  & 6999  & -109.778  & 4.719  & 942  \\[0.8ex]
        CO\textsubscript{2}                 & -188.180 & 9.308  & 11038 & -188.717  & 6.770  & 1191 \\[0.8ex]
        Li\textsubscript{2}                 & -15.126  & 1.508  & 4164  & -15.183   & 1.345  & 313  \\[0.8ex]
        CH\textsubscript{4}                 & -40.500  & 3.060  & 4306  & -40.691   & 2.606  & 681  \\[0.8ex]
        C\textsubscript{2}H\textsubscript{2}& -77.252  & 4.825  & 5837  & -77.498   & 3.812  & 1522 \\[0.8ex]
        C\textsubscript{2}H\textsubscript{4}& -78.466  & 4.916  & 7295  & -78.712   & 3.898  & 1146 \\[0.8ex]
        HCN                                 & -93.329  & 5.526  & 7254  & -93.634   & 4.276  & 2543 \\[0.8ex]
        NH\textsubscript{3}                 & -56.510  & 3.892  & 4385  & -56.789   & 3.176  & 2042 \\[0.8ex]
        H\textsubscript{2}CO                & -114.326 & 6.475  & 4981  & -114.731  & 4.924  & 942  \\[0.8ex]
        CH\textsubscript{3}OH               & -115.498 & 6.565  & 8653  & -115.902  & 5.004  & 1416 \\[0.8ex]
        H\textsubscript{2}O                 & -76.353  & 4.893  & 3671  & -76.749   & 3.829  & 1120 \\[0.8ex]
        H\textsubscript{2}O\textsubscript{2}& -151.260 & 8.013  & 5467  & -151.762  & 5.921  & 1367 \\[0.8ex]
        O\textsubscript{2}                  & -150.028 & 7.912  & 45765 & -150.543  & 5.815  & 2219 \\[0.8ex]
        HF                                  & -100.376 & 6.073 & 12371  & -100.846  & 4.687  & 1155 \\[0.8ex]
        \hline \hline
    \end{tabular}
}
\end{table}

We performed a series of calculations on 16 small organic molecules taken from the dataset of Ma \textit{et al} \cite{Ma2005} in order to
compare the performance of our AO-based scheme with existing MO-based
approaches.
As with all previous work to which we compare here, calculations were
performed without a Jastrow factor in order to isolate the impact of
the electron-nuclear cusp correction on the calculations.
In Table~\ref{table:normal_dat}, we report the average energy and its variance for both the uncusped and cusped cases, evaluating the
uncertainty in each by assuming normal statistics.
We should stress, however, that the absence of a Jastrow factor means
that electron-electron cusps are missing, and so the local energy
diverges when electrons approach each other closely.
This issue makes the quantities in Table~\ref{table:normal_dat} difficult
to interpret, as these divergences violate the central limit theorem
and lead to data that is not normally distributed.
Indeed, a standard quantile-quantile analysis of our data revealed
it to be non-normal with a significant rightward skew,
an effect that was even more pronounced in the uncusped data
where electron-nuclear divergences are also present.
Despite this issue, we begin with an analysis based on
the assumption of normal statistics in order to make an
apples-to-apples comparison to previous work.
After this comparison, we will present an analysis in which
we make no assumptions about normality.

In the first half of Table~\ref{table:normal_dat}, we compare our results for LiH, $\mathrm{C_2H_6}$, $\mathrm{N_2}$ and $\mathrm{CO_2}$ to Kussman \textit{et al.}\cite{Kussman2007}, Ma \textit{et al.}\cite{Ma2005}, and Per \textit{et al.}\cite{Per2008} (only for LiH and $\mathrm{N_2}$). Kussman \textit{et al.}\ use the same 6-31G(d) basis while Per \textit{et al.}\ use a 6-311G(d) basis. Ma \textit{et al.}\ do not specify which basis they use for their small-molecule analysis, but their reported HF energies suggest it is a larger basis than either 6-31G(d) or 6-311G(d).
These works also employed varying amounts of statistical sampling.
In this work, energy and variance uncertainties are evaluated using
5 million total samples by treating the results of ten independent
500,000-sample calculations as ten independent draws from an
unknown normal distribution.
Kussman \textit{et al.} used 500,000 total samples,
Ma \textit{et al.} used 250,000,
and Per \textit{et al.} left precise sample numbers unspecified.

Overall, all four approaches provide similar
variance reductions compared to uncusped results.
However, due to the use of different basis sets and
sampling efforts, and the fact that none of these
tests are expected to produce normally distributed data,
it is not possible to say with precision which schemes offer
larger variance reductions in which molecules.
For example, Kussman \textit{et al.}\ used the same basis set
and MP2/6-31G(d) reference geometries, yet the uncusped
variance uncertainties they report are at least
an order of magnitude smaller than ours.
This difference appears to be due to our larger sample size, which makes
our tests much more likely to encounter samples with near-divergent local
energies arising from electron-nuclear coalescence.
Although this issue makes uncertainties hard to interpret,
the AO-based cusping scheme nonetheless
shows very similar levels of variance reduction as the MO-based
approaches for the molecules that are shared in common,
and similarly large variance
reductions in the remaining molecules shown at the bottom
of Table~\ref{table:normal_dat}. \\
In Fig.~\ref{fig:El_v_MCMC}, we show detailed sampling data
for the same four molecules compared to other work in Table~\ref{table:normal_dat}  in order to highlight both the non-normality of
the data and the degree to which non-normality is reduced when
electron-nuclear cusps are added to the AOs.
First, we note that the vertical axis scale is in \textit{thousands}
of Hartree, which immediately makes clear that, as expected,
the data contains near-divergences due to particle coalescence
that are not consistent with a normal distribution.
Second, we see that the addition of nuclear cusps in the AOs
dramatically reduces these near-divergences, but that some
still remain due to electron-electron coalescence.
Thus, on the one hand, we are reassured that the addition of
electron-nuclear cusps is helping a great deal, but, on the
other hand, we would prefer to augment the discussion above
with measures of improvement that do not assume normally
distributed data.

To this purpose, Table~\ref{table:nonnormal_dat} reports the median,
interquartile range, and total range of the local energy across
all 5 million samples for each molecule in our test set.
In all cases, the local energy median and interquartile range
decrease when cusps are added to the AOs,
indicating both a variational improvement in the energy and a
reduction in the local energy's statistical spread that is
consistent with the variance analysis above.
As for the near near-divergences seen in Fig.~\ref{fig:El_v_MCMC},
these can be quantified via the total range measure.
By eliminating electron-nuclear divergences, the AO cusp correction
removes the most negative divergences, but it has little to no impact
on the positive divergences related to electron-electron coalescence.
Thus, the total range is significantly reduced by the AO
cusps but remains large, as seen in Table~\ref{table:nonnormal_dat}.
Nonetheless, these more general metrics all indicate that the
expected statistical improvements, which were tentatively indicated
by an analysis assuming normal statistics, are indeed real.

\section{Conclusion}

In conclusion, adding nuclear cusps directly to Gaussian atomic orbitals offers
similar advantages as molecular-orbital-based cusping methods while maintaining
independence from the many-body method that generates the molecular orbitals
and other trial function parameters.
In particular, we have shown that an interpolation between the original
Gaussian type orbital and a Slater-based function with a low-order polynomial
radial function produces electron-nuclear cusps that substantially improve the
statistics in VMC evaluations.
As with other cusping schemes, the update to the atomic orbitals can be
applied cheaply before starting any QMC sampling.
For convenience, we have provided a lightweight standalone library,
\textsc{CGAOWS},
that provides code both for determining the cusps and evaluating the
resulting cusped atomic orbitals.
All told, this approach provides all-electron VMC with local and transferrable
atomic orbital basis sets that, being based on those from quantum chemistry,
interface easily with quantum chemistry methods.

\section*{Supplementary Information}
See \href{run:SI.tex}{Supplementary Information} for electron positions used in the case studies from Section~\ref{sec:Ne} and a description of the \textsc{CGAOWS} software library.

\begin{acknowledgments}
This work was supported by the Office of Science, Office of Basic Energy Sciences, the U.S. Department of Energy, Contract Number DE-AC02-05CH11231, through the Gas Phase Chemical Physics program. Computational work was performed with the LBNL Lawrencium cluster and the Savio computational cluster resource provided by the Berkeley Research
Computing program at the University of California, Berkeley. T.K.Q. acknowledges that this material is based upon work supported by the National Science Foundation Graduate Research Fellowship Program under Grant No. DGE 2146752. Any opinions, findings, and conclusions or recommendations expressed in this material are those of the authors and do not necessarily reflect the views of the National Science Foundation.

\end{acknowledgments}

\section*{Data Availability Statement}

The data that support the findings of this study are available
within the article and its supplementary material.

%\renewcommand{\thesection}{\Roman{section}}
%\setcounter{section}{6}
\section*{References}
\bibliographystyle{achemso}
\bibliography{main}

\providecommand{\noopsort}[1]{}\providecommand{\singleletter}[1]{#1}%
\providecommand{\latin}[1]{#1}
\makeatletter
\providecommand{\doi}
  {\begingroup\let\do\@makeother\dospecials
  \catcode`\{=1 \catcode`\}=2 \doi@aux}
\providecommand{\doi@aux}[1]{\endgroup\texttt{#1}}
\makeatother
\providecommand*\mcitethebibliography{\thebibliography}
\csname @ifundefined\endcsname{endmcitethebibliography}  {\let\endmcitethebibliography\endthebibliography}{}
\begin{mcitethebibliography}{50}
\providecommand*\natexlab[1]{#1}
\providecommand*\mciteSetBstSublistMode[1]{}
\providecommand*\mciteSetBstMaxWidthForm[2]{}
\providecommand*\mciteBstWouldAddEndPuncttrue
  {\def\EndOfBibitem{\unskip.}}
\providecommand*\mciteBstWouldAddEndPunctfalse
  {\let\EndOfBibitem\relax}
\providecommand*\mciteSetBstMidEndSepPunct[3]{}
\providecommand*\mciteSetBstSublistLabelBeginEnd[3]{}
\providecommand*\EndOfBibitem{}
\mciteSetBstSublistMode{f}
\mciteSetBstMaxWidthForm{subitem}{(\alph{mcitesubitemcount})}
\mciteSetBstSublistLabelBeginEnd
  {\mcitemaxwidthsubitemform\space}
  {\relax}
  {\relax}

\bibitem[Dunning~Jr(1971)]{dunning1971}
Dunning~Jr,~T.~H. Gaussian basis functions for use in molecular calculations. IV. The representation of polarization functions for the first row atoms and hydrogen. \emph{The Journal of Chemical Physics} \textbf{1971}, \emph{55}, 3958--3966\relax
\mciteBstWouldAddEndPuncttrue
\mciteSetBstMidEndSepPunct{\mcitedefaultmidpunct}
{\mcitedefaultendpunct}{\mcitedefaultseppunct}\relax
\EndOfBibitem
\bibitem[Spitznagel \latin{et~al.}(1982)Spitznagel, Clark, Chandrasekhar, and Schleyer]{spitznagel1982}
Spitznagel,~G.~W.; Clark,~T.; Chandrasekhar,~J.; Schleyer,~P. V.~R. Stabilization of methyl anions by first-row substituents. The superiority of diffuse function-augmented basis sets for anion calculations. \emph{Journal of Computational Chemistry} \textbf{1982}, \emph{3}, 363--371\relax
\mciteBstWouldAddEndPuncttrue
\mciteSetBstMidEndSepPunct{\mcitedefaultmidpunct}
{\mcitedefaultendpunct}{\mcitedefaultseppunct}\relax
\EndOfBibitem
\bibitem[Kaufmann \latin{et~al.}(1989)Kaufmann, Baumeister, and Jungen]{kaufmann1989}
Kaufmann,~K.; Baumeister,~W.; Jungen,~M. Universal Gaussian basis sets for an optimum representation of Rydberg and continuum wavefunctions. \emph{Journal of Physics B: Atomic, Molecular and Optical Physics} \textbf{1989}, \emph{22}, 2223\relax
\mciteBstWouldAddEndPuncttrue
\mciteSetBstMidEndSepPunct{\mcitedefaultmidpunct}
{\mcitedefaultendpunct}{\mcitedefaultseppunct}\relax
\EndOfBibitem
\bibitem[Helgaker \latin{et~al.}(2013)Helgaker, J{\o}rgensen, and Olsen]{helgaker2013book}
Helgaker,~T.; J{\o}rgensen,~P.; Olsen,~J. \emph{Molecular Electronic-Structure Theory}; John Wiley \& Sons, 2013; pp 336--428\relax
\mciteBstWouldAddEndPuncttrue
\mciteSetBstMidEndSepPunct{\mcitedefaultmidpunct}
{\mcitedefaultendpunct}{\mcitedefaultseppunct}\relax
\EndOfBibitem
\bibitem[Foulkes \latin{et~al.}(2001)Foulkes, Mitas, Needs, and Rajagopal]{Foulkes}
Foulkes,~W. M.~C.; Mitas,~L.; Needs,~R.~J.; Rajagopal,~G. Quantum Monte Carlo simulations of solids. \emph{Rev. Mod. Phys.} \textbf{2001}, \emph{73}, 33--83\relax
\mciteBstWouldAddEndPuncttrue
\mciteSetBstMidEndSepPunct{\mcitedefaultmidpunct}
{\mcitedefaultendpunct}{\mcitedefaultseppunct}\relax
\EndOfBibitem
\bibitem[Otis and Neuscamman(2023)Otis, and Neuscamman]{LeonRev}
Otis,~L.; Neuscamman,~E. A promising intersection of excited-state-specific methods from quantum chemistry and quantum Monte Carlo. \emph{WIREs Computational Molecular Science} \textbf{2023}, \emph{13}, e1659\relax
\mciteBstWouldAddEndPuncttrue
\mciteSetBstMidEndSepPunct{\mcitedefaultmidpunct}
{\mcitedefaultendpunct}{\mcitedefaultseppunct}\relax
\EndOfBibitem
\bibitem[Shepard \latin{et~al.}(2022)Shepard, Panadés-Barrueta, Moroni, Scemama, and Filippi]{Shepard}
Shepard,~S.; Panadés-Barrueta,~R.~L.; Moroni,~S.; Scemama,~A.; Filippi,~C. Double Excitation Energies from Quantum Monte Carlo Using State-Specific Energy Optimization. \emph{Journal of Chemical Theory and Computation} \textbf{2022}, \emph{18}, 6722--6731\relax
\mciteBstWouldAddEndPuncttrue
\mciteSetBstMidEndSepPunct{\mcitedefaultmidpunct}
{\mcitedefaultendpunct}{\mcitedefaultseppunct}\relax
\EndOfBibitem
\bibitem[Garner and Neuscamman(2020)Garner, and Neuscamman]{Scott2020}
Garner,~S.~M.; Neuscamman,~E. {A variational Monte Carlo approach for core excitations}. \emph{The Journal of Chemical Physics} \textbf{2020}, \emph{153}, 144108\relax
\mciteBstWouldAddEndPuncttrue
\mciteSetBstMidEndSepPunct{\mcitedefaultmidpunct}
{\mcitedefaultendpunct}{\mcitedefaultseppunct}\relax
\EndOfBibitem
\bibitem[Dash \latin{et~al.}(2021)Dash, Moroni, Filippi, and Scemama]{Dash2021}
Dash,~M.; Moroni,~S.; Filippi,~C.; Scemama,~A. Tailoring CIPSI Expansions for QMC Calculations of Electronic Excitations: The Case Study of Thiophene. \emph{Journal of Chemical Theory and Computation} \textbf{2021}, \emph{17}, 3426--3434\relax
\mciteBstWouldAddEndPuncttrue
\mciteSetBstMidEndSepPunct{\mcitedefaultmidpunct}
{\mcitedefaultendpunct}{\mcitedefaultseppunct}\relax
\EndOfBibitem
\bibitem[Ludovicy \latin{et~al.}(2023)Ludovicy, Dahl, and Luchow]{Ludovicy2023}
Ludovicy,~J.; Dahl,~R.; Luchow,~A. Toward Compact Selected Configuration Interaction Wave Functions with Quantum Monte Carlo─ A Case Study of C2. \emph{Journal of Chemical Theory and Computation} \textbf{2023}, \emph{19}, 2792--2803\relax
\mciteBstWouldAddEndPuncttrue
\mciteSetBstMidEndSepPunct{\mcitedefaultmidpunct}
{\mcitedefaultendpunct}{\mcitedefaultseppunct}\relax
\EndOfBibitem
\bibitem[Clark \latin{et~al.}(2011)Clark, Morales, McMinis, Kim, and Scuseria]{Clark2011}
Clark,~B.~K.; Morales,~M.~A.; McMinis,~J.; Kim,~J.; Scuseria,~G.~E. {Computing the energy of a water molecule using multideterminants: A simple, efficient algorithm}. \emph{The Journal of Chemical Physics} \textbf{2011}, \emph{135}, 244105\relax
\mciteBstWouldAddEndPuncttrue
\mciteSetBstMidEndSepPunct{\mcitedefaultmidpunct}
{\mcitedefaultendpunct}{\mcitedefaultseppunct}\relax
\EndOfBibitem
\bibitem[Assaraf \latin{et~al.}(2017)Assaraf, Moroni, and Filippi]{Assaraf2017}
Assaraf,~R.; Moroni,~S.; Filippi,~C. Optimizing the Energy with Quantum Monte Carlo: A Lower Numerical Scaling for Jastrow–Slater Expansions. \emph{Journal of Chemical Theory and Computation} \textbf{2017}, \emph{13}, 5273--5281\relax
\mciteBstWouldAddEndPuncttrue
\mciteSetBstMidEndSepPunct{\mcitedefaultmidpunct}
{\mcitedefaultendpunct}{\mcitedefaultseppunct}\relax
\EndOfBibitem
\bibitem[Petruzielo \latin{et~al.}(2012)Petruzielo, Toulouse, and Umrigar]{Petruzielo}
Petruzielo,~F.~R.; Toulouse,~J.; Umrigar,~C.~J. {Approaching chemical accuracy with quantum Monte Carlo}. \emph{The Journal of Chemical Physics} \textbf{2012}, \emph{136}, 124116\relax
\mciteBstWouldAddEndPuncttrue
\mciteSetBstMidEndSepPunct{\mcitedefaultmidpunct}
{\mcitedefaultendpunct}{\mcitedefaultseppunct}\relax
\EndOfBibitem
\bibitem[Emmanuel~Giner and Toulouse(2016)Emmanuel~Giner, and Toulouse]{Giner2016}
Emmanuel~Giner,~R.~A.; Toulouse,~J. Quantum Monte Carlo with reoptimised perturbatively selected configuration-interaction wave functions. \emph{Molecular Physics} \textbf{2016}, \emph{114}, 910--920\relax
\mciteBstWouldAddEndPuncttrue
\mciteSetBstMidEndSepPunct{\mcitedefaultmidpunct}
{\mcitedefaultendpunct}{\mcitedefaultseppunct}\relax
\EndOfBibitem
\bibitem[Dash \latin{et~al.}(2019)Dash, Feldt, Moroni, Scemama, and Filippi]{Dash2019}
Dash,~M.; Feldt,~J.; Moroni,~S.; Scemama,~A.; Filippi,~C. Excited States with Selected Configuration Interaction-Quantum Monte Carlo: Chemically Accurate Excitation Energies and Geometries. \emph{Journal of Chemical Theory and Computation} \textbf{2019}, \emph{15}, 4896--4906\relax
\mciteBstWouldAddEndPuncttrue
\mciteSetBstMidEndSepPunct{\mcitedefaultmidpunct}
{\mcitedefaultendpunct}{\mcitedefaultseppunct}\relax
\EndOfBibitem
\bibitem[Pineda~Flores and Neuscamman(2019)Pineda~Flores, and Neuscamman]{Sergio2019}
Pineda~Flores,~S.~D.; Neuscamman,~E. Excited State Specific Multi-Slater Jastrow Wave Functions. \emph{The Journal of Physical Chemistry A} \textbf{2019}, \emph{123}, 1487--1497\relax
\mciteBstWouldAddEndPuncttrue
\mciteSetBstMidEndSepPunct{\mcitedefaultmidpunct}
{\mcitedefaultendpunct}{\mcitedefaultseppunct}\relax
\EndOfBibitem
\bibitem[Otis \latin{et~al.}(2020)Otis, Craig, and Neuscamman]{Leon2020}
Otis,~L.; Craig,~I.~M.; Neuscamman,~E. {A hybrid approach to excited-state-specific variational Monte Carlo and doubly excited states}. \emph{The Journal of Chemical Physics} \textbf{2020}, \emph{153}, 234105\relax
\mciteBstWouldAddEndPuncttrue
\mciteSetBstMidEndSepPunct{\mcitedefaultmidpunct}
{\mcitedefaultendpunct}{\mcitedefaultseppunct}\relax
\EndOfBibitem
\bibitem[Pfau \latin{et~al.}(2020)Pfau, Spencer, Matthews, and Foulkes]{Ferminet2020}
Pfau,~D.; Spencer,~J.~S.; Matthews,~A. G. D.~G.; Foulkes,~W. M.~C. Ab initio solution of the many-electron Schr\"odinger equation with deep neural networks. \emph{Phys. Rev. Res.} \textbf{2020}, \emph{2}, 033429\relax
\mciteBstWouldAddEndPuncttrue
\mciteSetBstMidEndSepPunct{\mcitedefaultmidpunct}
{\mcitedefaultendpunct}{\mcitedefaultseppunct}\relax
\EndOfBibitem
\bibitem[Hermann \latin{et~al.}(2020)Hermann, Schätzle, and Noé]{Paulinet2020}
Hermann,~J.; Schätzle,~Z.; Noé,~F. Deep-neural-network solution of the electronic Schrödinger equation. \emph{Nat. Chem.} \textbf{2020}, \emph{12}, 891–897\relax
\mciteBstWouldAddEndPuncttrue
\mciteSetBstMidEndSepPunct{\mcitedefaultmidpunct}
{\mcitedefaultendpunct}{\mcitedefaultseppunct}\relax
\EndOfBibitem
\bibitem[Hermann \latin{et~al.}(2023)Hermann, Spencer, Choo, Mezzacapo, Foulkes, Pfau, Carleo, and Noé]{MLrev2023}
Hermann,~J.; Spencer,~J.; Choo,~K.; Mezzacapo,~A.; Foulkes,~W. M.~C.; Pfau,~D.; Carleo,~G.; Noé,~F. Ab initio quantum chemistry with neural-network wavefunctions. \emph{Nat. Rev. Chem.} \textbf{2023}, \emph{7}, 692–709\relax
\mciteBstWouldAddEndPuncttrue
\mciteSetBstMidEndSepPunct{\mcitedefaultmidpunct}
{\mcitedefaultendpunct}{\mcitedefaultseppunct}\relax
\EndOfBibitem
\bibitem[Nicklass \latin{et~al.}(1995)Nicklass, Dolg, Stoll, and Preuss]{STU}
Nicklass,~A.; Dolg,~M.; Stoll,~H.; Preuss,~H. {Ab initio energy‐adjusted pseudopotentials for the noble gases Ne through Xe: Calculation of atomic dipole and quadrupole polarizabilities}. \emph{The Journal of Chemical Physics} \textbf{1995}, \emph{102}, 8942--8952\relax
\mciteBstWouldAddEndPuncttrue
\mciteSetBstMidEndSepPunct{\mcitedefaultmidpunct}
{\mcitedefaultendpunct}{\mcitedefaultseppunct}\relax
\EndOfBibitem
\bibitem[Burkatzki \latin{et~al.}(2007)Burkatzki, Filippi, and Dolg]{BFD}
Burkatzki,~M.; Filippi,~C.; Dolg,~M. {Energy-consistent pseudopotentials for quantum Monte Carlo calculations}. \emph{The Journal of Chemical Physics} \textbf{2007}, \emph{126}, 234105\relax
\mciteBstWouldAddEndPuncttrue
\mciteSetBstMidEndSepPunct{\mcitedefaultmidpunct}
{\mcitedefaultendpunct}{\mcitedefaultseppunct}\relax
\EndOfBibitem
\bibitem[Trail and Needs(2017)Trail, and Needs]{cECP}
Trail,~J.~R.; Needs,~R.~J. {Shape and energy consistent pseudopotentials for correlated electron systems}. \emph{The Journal of Chemical Physics} \textbf{2017}, \emph{146}, 204107\relax
\mciteBstWouldAddEndPuncttrue
\mciteSetBstMidEndSepPunct{\mcitedefaultmidpunct}
{\mcitedefaultendpunct}{\mcitedefaultseppunct}\relax
\EndOfBibitem
\bibitem[Wang \latin{et~al.}(2022)Wang, Kincaid, Zhou, Annaberdiyev, Bennett, Krogel, and Mitas]{ccECP}
Wang,~G.; Kincaid,~B.; Zhou,~H.; Annaberdiyev,~A.; Bennett,~M.~C.; Krogel,~J.~T.; Mitas,~L. {A new generation of effective core potentials from correlated and spin–orbit calculations: Selected heavy elements}. \emph{The Journal of Chemical Physics} \textbf{2022}, \emph{157}, 054101\relax
\mciteBstWouldAddEndPuncttrue
\mciteSetBstMidEndSepPunct{\mcitedefaultmidpunct}
{\mcitedefaultendpunct}{\mcitedefaultseppunct}\relax
\EndOfBibitem
\bibitem[Dolg and Cao(2012)Dolg, and Cao]{dolg2012relativistic}
Dolg,~M.; Cao,~X. Relativistic pseudopotentials: their development and scope of applications. \emph{Chemical reviews} \textbf{2012}, \emph{112}, 403--480\relax
\mciteBstWouldAddEndPuncttrue
\mciteSetBstMidEndSepPunct{\mcitedefaultmidpunct}
{\mcitedefaultendpunct}{\mcitedefaultseppunct}\relax
\EndOfBibitem
\bibitem[Wang \latin{et~al.}(2024)Wang, Zhou, and Wang]{MLECP}
Wang,~M.; Zhou,~Y.; Wang,~H. {Performance assessment of the effective core potentials under the fermionic neural network: First and second row elements}. \emph{The Journal of Chemical Physics} \textbf{2024}, \emph{160}, 204109\relax
\mciteBstWouldAddEndPuncttrue
\mciteSetBstMidEndSepPunct{\mcitedefaultmidpunct}
{\mcitedefaultendpunct}{\mcitedefaultseppunct}\relax
\EndOfBibitem
\bibitem[Casula(2006)]{Casula2006}
Casula,~M. Beyond the locality approximation in the standard diffusion Monte Carlo method. \emph{Phys. Rev. B} \textbf{2006}, \emph{74}, 161102\relax
\mciteBstWouldAddEndPuncttrue
\mciteSetBstMidEndSepPunct{\mcitedefaultmidpunct}
{\mcitedefaultendpunct}{\mcitedefaultseppunct}\relax
\EndOfBibitem
\bibitem[Krogel and Kent(2017)Krogel, and Kent]{Krogel}
Krogel,~J.~T.; Kent,~P. R.~C. {Magnitude of pseudopotential localization errors in fixed node diffusion quantum Monte Carlo}. \emph{The Journal of Chemical Physics} \textbf{2017}, \emph{146}, 244101\relax
\mciteBstWouldAddEndPuncttrue
\mciteSetBstMidEndSepPunct{\mcitedefaultmidpunct}
{\mcitedefaultendpunct}{\mcitedefaultseppunct}\relax
\EndOfBibitem
\bibitem[Scemama \latin{et~al.}(2019)Scemama, Caffarel, Benali, Jacquemin, and Loos]{Scemama2019}
Scemama,~A.; Caffarel,~M.; Benali,~A.; Jacquemin,~D.; Loos,~P.-F. Influence of pseudopotentials on excitation energies from selected configuration interaction and diffusion Monte Carlo. \emph{Results in Chemistry} \textbf{2019}, \emph{1}, 100002\relax
\mciteBstWouldAddEndPuncttrue
\mciteSetBstMidEndSepPunct{\mcitedefaultmidpunct}
{\mcitedefaultendpunct}{\mcitedefaultseppunct}\relax
\EndOfBibitem
\bibitem[Kato(1957)]{katoEigenfunctionsManyparticleSystems1957a}
Kato,~T. On the Eigenfunctions of Many-Particle Systems in Quantum Mechanics. \emph{Communications on Pure and Applied Mathematics} \textbf{1957}, \emph{10}, 151--177\relax
\mciteBstWouldAddEndPuncttrue
\mciteSetBstMidEndSepPunct{\mcitedefaultmidpunct}
{\mcitedefaultendpunct}{\mcitedefaultseppunct}\relax
\EndOfBibitem
\bibitem[Drummond \latin{et~al.}(2004)Drummond, Towler, and Needs]{Drummond2004}
Drummond,~N.~D.; Towler,~M.~D.; Needs,~R.~J. Jastrow correlation factor for atoms, molecules, and solids. \emph{Phys. Rev. B} \textbf{2004}, \emph{70}, 235119\relax
\mciteBstWouldAddEndPuncttrue
\mciteSetBstMidEndSepPunct{\mcitedefaultmidpunct}
{\mcitedefaultendpunct}{\mcitedefaultseppunct}\relax
\EndOfBibitem
\bibitem[Slater(1930)]{STO}
Slater,~J.~C. Atomic Shielding Constants. \emph{Phys. Rev.} \textbf{1930}, \emph{36}, 57--64\relax
\mciteBstWouldAddEndPuncttrue
\mciteSetBstMidEndSepPunct{\mcitedefaultmidpunct}
{\mcitedefaultendpunct}{\mcitedefaultseppunct}\relax
\EndOfBibitem
\bibitem[te~Velde \latin{et~al.}(2001)te~Velde, Bickelhaupt, Baerends, Fonseca~Guerra, van Gisbergen, Snijders, and Ziegler]{AMS2001}
te~Velde,~G.; Bickelhaupt,~F.~M.; Baerends,~E.~J.; Fonseca~Guerra,~C.; van Gisbergen,~S. J.~A.; Snijders,~J.~G.; Ziegler,~T. Chemistry with ADF. \emph{Journal of Computational Chemistry} \textbf{2001}, \emph{22}, 931--967\relax
\mciteBstWouldAddEndPuncttrue
\mciteSetBstMidEndSepPunct{\mcitedefaultmidpunct}
{\mcitedefaultendpunct}{\mcitedefaultseppunct}\relax
\EndOfBibitem
\bibitem[Ma \latin{et~al.}(2005)Ma, Towler, Drummond, and Needs]{Ma2005}
Ma,~A.; Towler,~M.~D.; Drummond,~N.~D.; Needs,~R.~J. {Scheme for adding electron–nucleus cusps to Gaussian orbitals}. \emph{J. Chem. Phys.} \textbf{2005}, \emph{122}, 224322\relax
\mciteBstWouldAddEndPuncttrue
\mciteSetBstMidEndSepPunct{\mcitedefaultmidpunct}
{\mcitedefaultendpunct}{\mcitedefaultseppunct}\relax
\EndOfBibitem
\bibitem[Per \latin{et~al.}(2008)Per, Russo, and Snook]{Per2008}
Per,~M.~C.; Russo,~S.~P.; Snook,~I.~K. Electron-Nucleus Cusp Correction and Forces in Quantum {{Monte Carlo}}. \emph{The Journal of Chemical Physics} \textbf{2008}, \emph{128}, 114106\relax
\mciteBstWouldAddEndPuncttrue
\mciteSetBstMidEndSepPunct{\mcitedefaultmidpunct}
{\mcitedefaultendpunct}{\mcitedefaultseppunct}\relax
\EndOfBibitem
\bibitem[Manten and Lüchow(2001)Manten, and Lüchow]{Manten2001}
Manten,~S.; Lüchow,~A. On the Accuracy of the Fixed-Node Diffusion Quantum {{Monte Carlo}} Method. \emph{The Journal of Chemical Physics} \textbf{2001}, \emph{115}, 5362--5366\relax
\mciteBstWouldAddEndPuncttrue
\mciteSetBstMidEndSepPunct{\mcitedefaultmidpunct}
{\mcitedefaultendpunct}{\mcitedefaultseppunct}\relax
\EndOfBibitem
\bibitem[Loos \latin{et~al.}(2019)Loos, Scemama, and Caffarel]{Loos2019}
Loos,~P.-F.; Scemama,~A.; Caffarel,~M. In \emph{Advances in {{Quantum Chemistry}}}; Ancarani,~L.~U., Hoggan,~P.~E., Eds.; State of {{The Art}} of {{Molecular Electronic Structure Computations}}: {{Correlation Methods}}, {{Basis Sets}} and {{More}}; Academic Press, 2019; Vol.~79; pp 113--132\relax
\mciteBstWouldAddEndPuncttrue
\mciteSetBstMidEndSepPunct{\mcitedefaultmidpunct}
{\mcitedefaultendpunct}{\mcitedefaultseppunct}\relax
\EndOfBibitem
\bibitem[Nakatsuka \latin{et~al.}(2010)Nakatsuka, Nakajima, and Hirao]{nakatsuka}
Nakatsuka,~Y.; Nakajima,~T.; Hirao,~K. Electron-Nucleus Cusp Correction Scheme for the Relativistic Zeroth-Order Regular Approximation Quantum {{Monte Carlo}} Method. \emph{The Journal of Chemical Physics} \textbf{2010}, \emph{132}, 174108\relax
\mciteBstWouldAddEndPuncttrue
\mciteSetBstMidEndSepPunct{\mcitedefaultmidpunct}
{\mcitedefaultendpunct}{\mcitedefaultseppunct}\relax
\EndOfBibitem
\bibitem[Kussmann and Ochsenfeld(2007)Kussmann, and Ochsenfeld]{Kussman2007}
Kussmann,~J.; Ochsenfeld,~C. Adding Electron-Nuclear Cusps to {{Gaussian}} Basis Functions for Molecular Quantum {{Monte Carlo}} Calculations. \emph{Physical Review B} \textbf{2007}, \emph{76}, 115115\relax
\mciteBstWouldAddEndPuncttrue
\mciteSetBstMidEndSepPunct{\mcitedefaultmidpunct}
{\mcitedefaultendpunct}{\mcitedefaultseppunct}\relax
\EndOfBibitem
\bibitem[Broyden(1970)]{Broyden}
Broyden,~C.~G. The convergence of a class of double-rankminimization algorithms 1. general considerations. \emph{IMA J. Appl. Math.} \textbf{1970}, \emph{6}, 76--−90\relax
\mciteBstWouldAddEndPuncttrue
\mciteSetBstMidEndSepPunct{\mcitedefaultmidpunct}
{\mcitedefaultendpunct}{\mcitedefaultseppunct}\relax
\EndOfBibitem
\bibitem[Fletcher(1970)]{Fletcher}
Fletcher,~R. A new approach to variable metric algorithms. \emph{J.Comput.} \textbf{1970}, \emph{13}, 317--−322\relax
\mciteBstWouldAddEndPuncttrue
\mciteSetBstMidEndSepPunct{\mcitedefaultmidpunct}
{\mcitedefaultendpunct}{\mcitedefaultseppunct}\relax
\EndOfBibitem
\bibitem[Goldfarb(1970)]{Goldfarb}
Goldfarb,~D. A family of variable-metric methods derived by variational means. \emph{Math. Comput.} \textbf{1970}, \emph{24}, 23--26\relax
\mciteBstWouldAddEndPuncttrue
\mciteSetBstMidEndSepPunct{\mcitedefaultmidpunct}
{\mcitedefaultendpunct}{\mcitedefaultseppunct}\relax
\EndOfBibitem
\bibitem[Shanno(1970)]{Shanno}
Shanno,~D.~F. Conditioning of quasi-Newton methods forfunction minimization. \emph{Math. Comput.} \textbf{1970}, \emph{24}, 647−--656\relax
\mciteBstWouldAddEndPuncttrue
\mciteSetBstMidEndSepPunct{\mcitedefaultmidpunct}
{\mcitedefaultendpunct}{\mcitedefaultseppunct}\relax
\EndOfBibitem
\bibitem[Assaraf and Caffarel(1999)Assaraf, and Caffarel]{assarafZeroVariancePrincipleMonte1999}
Assaraf,~R.; Caffarel,~M. Zero-{{Variance Principle}} for {{Monte Carlo Algorithms}}. \emph{Physical Review Letters} \textbf{1999}, \emph{83}, 4682--4685\relax
\mciteBstWouldAddEndPuncttrue
\mciteSetBstMidEndSepPunct{\mcitedefaultmidpunct}
{\mcitedefaultendpunct}{\mcitedefaultseppunct}\relax
\EndOfBibitem
\bibitem[Foster and Boys(1960)Foster, and Boys]{fosterCanonicalConfigurationalInteraction1960}
Foster,~J.~M.; Boys,~S.~F. Canonical {{Configurational Interaction Procedure}}. \emph{Reviews of Modern Physics} \textbf{1960}, \emph{32}, 300--302\relax
\mciteBstWouldAddEndPuncttrue
\mciteSetBstMidEndSepPunct{\mcitedefaultmidpunct}
{\mcitedefaultendpunct}{\mcitedefaultseppunct}\relax
\EndOfBibitem
\bibitem[Sun(2015)]{sunLibcintEfficientGeneral2015}
Sun,~Q. Libcint: {{An}} Efficient General Integral Library for {{Gaussian}} Basis Functions. \emph{Journal of Computational Chemistry} \textbf{2015}, \emph{36}, 1664--1671\relax
\mciteBstWouldAddEndPuncttrue
\mciteSetBstMidEndSepPunct{\mcitedefaultmidpunct}
{\mcitedefaultendpunct}{\mcitedefaultseppunct}\relax
\EndOfBibitem
\bibitem[Sun \latin{et~al.}(2018)Sun, Berkelbach, Blunt, Booth, Guo, Li, Liu, McClain, Sayfutyarova, Sharma, Wouters, and Chan]{sunPySCFPythonbasedSimulations2018}
Sun,~Q.; Berkelbach,~T.~C.; Blunt,~N.~S.; Booth,~G.~H.; Guo,~S.; Li,~Z.; Liu,~J.; McClain,~J.~D.; Sayfutyarova,~E.~R.; Sharma,~S.; Wouters,~S.; Chan,~G. K.-L. {{PySCF}}: The {{Python-based}} Simulations of Chemistry Framework. \emph{WIREs Computational Molecular Science} \textbf{2018}, \emph{8}, e1340\relax
\mciteBstWouldAddEndPuncttrue
\mciteSetBstMidEndSepPunct{\mcitedefaultmidpunct}
{\mcitedefaultendpunct}{\mcitedefaultseppunct}\relax
\EndOfBibitem
\bibitem[Sun \latin{et~al.}(2020)Sun, Zhang, Banerjee, Bao, Barbry, Blunt, Bogdanov, Booth, Chen, Cui, Eriksen, Gao, Guo, Hermann, Hermes, Koh, Koval, Lehtola, Li, Liu, Mardirossian, McClain, Motta, Mussard, Pham, Pulkin, Purwanto, Robinson, Ronca, Sayfutyarova, Scheurer, Schurkus, Smith, Sun, Sun, Upadhyay, Wagner, Wang, White, Whitfield, Williamson, Wouters, Yang, Yu, Zhu, Berkelbach, Sharma, Sokolov, and Chan]{sunRecentDevelopmentsPySCF2020}
Sun,~Q. \latin{et~al.}  Recent Developments in the {{PySCF}} Program Package. \emph{The Journal of Chemical Physics} \textbf{2020}, \emph{153}, 024109\relax
\mciteBstWouldAddEndPuncttrue
\mciteSetBstMidEndSepPunct{\mcitedefaultmidpunct}
{\mcitedefaultendpunct}{\mcitedefaultseppunct}\relax
\EndOfBibitem
\bibitem[Hehre \latin{et~al.}(1972)Hehre, Ditchfield, and Pople]{Pople}
Hehre,~W.~J.; Ditchfield,~R.; Pople,~J.~A. \emph{J. Chem. Phys.} \textbf{1972}, \emph{56}, 2257\relax
\mciteBstWouldAddEndPuncttrue
\mciteSetBstMidEndSepPunct{\mcitedefaultmidpunct}
{\mcitedefaultendpunct}{\mcitedefaultseppunct}\relax
\EndOfBibitem
\bibitem[III(2022)]{cccdbd}
III,~R. D.~J. NIST Computational Chemistry Comparison and Benchmark Database, NIST Standard Reference Database Number 101. \url{http://cccbdb.nist.gov/}, 2022; \url{http://cccbdb.nist.gov/}\relax
\mciteBstWouldAddEndPuncttrue
\mciteSetBstMidEndSepPunct{\mcitedefaultmidpunct}
{\mcitedefaultendpunct}{\mcitedefaultseppunct}\relax
\EndOfBibitem
\end{mcitethebibliography}

\clearpage
\onecolumngrid
\section{Supplementary Information}
\renewcommand{\thesection}{S\arabic{section}}
\renewcommand{\theequation}{S\arabic{equation}}
\renewcommand{\thefigure}{S\arabic{figure}}
\renewcommand{\thetable}{S\arabic{table}}
\setcounter{section}{0}
\setcounter{figure}{0}
\setcounter{equation}{0}
\setcounter{table}{0}
\subsection{Electron Positions and Geometries for Nuclear Walk-Through Tests}
%% Add to SI potentially:
%   The electron configuration for both test cases (neon and methanol) were determined by normally distributing (standard deviation of $0.2$ Bohr) each electron around the nucleus (or middle of bond) the electron would inhabit in a Lewis diagram of the system. The alpha electron closest to the desired nucleus was chosen to ``walk'' through the nucleus. 
%\large
\begin{table}[h]
%\footnotesize
\centering
\caption{Coordinates of the frozen electrons in the neon walk-through test (see Figure 3 in the paper). The nucleus is centered at the origin. A and B denote alpha and beta spin, respectively. A1 corresponds to the electron being walked through the nucleus. Coordinates are in Bohr.
} 
%\resizebox{\columnwidth}{!}{%
\begin{tabular}{c S[table-format=3.6] S[table-format=3.6] S[table-format=3.6]}
\hline \hline
Electron & x & y & z \\
\hline
A1 & \text{--} & \text{--} & \text{--} \\

A2 & 0.213553 & -0.444243 & 0.278260 \\

A3 & -0.494929 & 0.584347 & -0.265310 \\

A4 & 0.372814 & 0.868481 & -1.627833 \\

A5 & -0.900304 & -1.472222 & -1.127845 \\

B1 & -0.919444 & -0.636217 & -0.295044 \\

B2 & 1.001112 & 0.185600 & -0.238352 \\

B3 & -0.598423 & -0.200966 & 0.797388 \\

B4 & -0.274741 & 0.465994 & -0.980761 \\

B5 & 0.062476 & -0.088619 & -0.033220 \\
\hline \hline
\end{tabular}
%}
\end{table}

\begin{table}[h]
%\footnotesize
\centering
\caption{Molecular geometry of methanol for walk-through test (see Figure 4 in the paper). Coordinates are in Bohr.
} 
%\resizebox{\columnwidth}{!}{%
\begin{tabular}{c S[table-format=3.6] S[table-format=3.6] S[table-format=3.6]}
\hline \hline
Z & x & y & z \\
\hline
C & -0.088460 & 1.259762 & 0.000000 \\    
O & -0.088460 & -1.434354 & 0.000000 \\     
H & -2.060765 & 1.850028 & 0.000000    \\    
H &  0.829924 & 2.030361 & 1.686264 \\
H &  0.829924 & 2.030361 & -1.686264 \\
H &  1.639359 &-1.994492 & 0.000000 \\

\hline \hline
\end{tabular}
%}
\end{table}

\begin{table}[htb]
%\footnotesize
\centering
\caption{Coordinates of the frozen electrons in the methanol walk-through test (see Figure 4 in the paper). A and B denote alpha and beta spin, respectively. A1 corresponds to the electron being walked through the carbon nucleus. Coordinates are in Bohr.
} 
%\resizebox{\columnwidth}{!}{%
\begin{tabular}{c S[table-format=3.6] S[table-format=3.6] S[table-format=3.6]}
\hline \hline
Electron & x & y & z \\
\hline
A1 & \text{--} & \text{--} & \text{--} \\

A2 & 0.106702 & 0.914076 & 0.381221  \\
A3 & 0.763427 & 1.029360 & 0.300633 \\
A4 & -1.214422 & 1.824894 &-0.535917  \\
A5 &-1.091921 & 1.072275  &1.074244 \\
A6 &-0.398179 &-1.133658 &-0.590961 \\
A7 &-0.095003 &-1.563129 &-0.045628 \\
A8 &-0.126748 &-1.720242 & 0.257205  \\
A9 & 1.161947 &-2.956023 &-0.126195\\

B1 &-0.222989 &-0.861524 &-0.514105  \\
B2 & 1.083038 &-0.238063 & 1.820019\\
B3 &-0.179891 &-1.328974 & 0.116196 \\
B4 & 0.653737 &-2.122696 &-0.284307 \\
B5 &-1.353428 & 2.401788 & 1.105960 \\
B6 &-1.211755 &-2.277452 &-0.687876 \\
B7 & 0.697330 & 1.929689 & 1.448539 \\
B8 & 0.063754 & 1.337925 &-0.142632 \\
B9 &-0.084558 & 0.904812 &-1.026934\\

\hline \hline
\end{tabular}
%}
\end{table}

\normalsize
\subsection{Cusping Gaussian Atomic Orbitals with Slaters (CGAOWS) Package}

We provide an open-source C++ library that produces cusped atomic orbitals for a chosen molecular system and basis set (currently limited to STO-3G, 6-31G, and 6-31G(d)). It can be found here: \hyperlink{https://github.com/eneuscamman/cgaows}{https://github.com/eneuscamman/cgaows}. After the user specifies the molecular details in the input file, Python code is used to calculate the cusp parameters and molecular orbital coefficient matrix that takes into account s-type orbital orthogonalization. These parameters are printed into txt files that are subsequently read into the C++ code. We provide functions that evaluate the AOs for a given electron configuration as well as the first and second orbital derivatives (cartesian) required for the evaluation of the kinetic energy. We include an example (Ne atom/6-31G(d) basis set) of how the package can be integrated into a variational Monte Carlo code. See the code for further details.

\end{document}